\documentclass[twocolumn,showpacs,preprintnumbers,amsmath,amssymb,pra,aps,
    superscriptaddress,longbibliography]{revtex4-1}
\usepackage{graphicx}
\usepackage{multirow}
\usepackage{amsmath}
\usepackage{soul}
\usepackage{multirow}
\usepackage{verbatim}
\usepackage{bm}
\usepackage{outlines}
\usepackage{siunitx}
\usepackage[space]{grffile}
\usepackage[colorlinks=true, linkcolor=blue, citecolor=gray]{hyperref}
\usepackage[capitalize]{cleveref}
\usepackage{xr}
\usepackage{color}
\usepackage{tikz}
\usepackage{mathtools}
\usepackage{amssymb}
\usetikzlibrary{calc, positioning}

\newcommand{\ket}[1]{\left\lvert #1 \right\rangle}
\newcommand{\Qzero}{Q_{0}}
\newcommand{\Qone}{Q_{1}}
\newcommand{\Qtwo}{Q_{2}}
\newcommand{\Rzero}{R_0}
\newcommand{\Rone}{R_1}
\newcommand{\ns}{\mathrm{ns}}

\newcommand{\us}{\mu\mathrm{s}}
\newcommand{\V}{\mathrm{V}}

\newcommand{\GHz}{\mathrm{GHz}}
\newcommand{\nm}{\mathrm{nm}}
\newcommand{\um}{\mu \mathrm{m}}
\newcommand{\mm}{\mathrm{mm}}
\newcommand{\mK}{\mathrm{mK}}
\newcommand{\dB}{\mathrm{dB}}
\newcommand{\K}{\mathrm{K}}
\newcommand{\Htwo}{\mathrm{H}_2}
\newcommand{\ham}{\hat{H}}
\newcommand{\hamHtwo}{\ham_{\mathrm{H}2}}
\newcommand{\iSWAP}{\textrm{iSWAP}}
\newcommand{\pauligroup}{\mathbb{P}}
\newcommand{\trace}{\mathrm{Tr}}
\newcommand{\rhoexpt}{\rho^{(\mathrm{raw})}}
\newcommand{\rhosim}{\rho^{(\mathrm{sim})}}
\newcommand{\rhosimp}{\rho_{\hat{P}}^{(\mathrm{sim})}}

\newcommand{\rhosv}{\rho^{(\mathrm{SV})}}
\newcommand{\rhotrue}{\rho^{(0)}}

\newcommand{\Eraw}{E^{(\mathrm{raw})}}
\newcommand{\Esv}{E^{(\mathrm{SV})}}
\newcommand{\iSWAPtheta}{\textrm{U}_{\theta}}
\newcommand{\Fidel}{F}

\newcommand{\Eerr}{\Delta E}
\newcommand{\Nmeas}{N_{\mathrm{meas}}}
\newcommand{\Tone}{T_{1}}
\newcommand{\Ttwos}{T_{2}^*}

\newcommand{\Ttwosred}{T_2^{*,\textrm{red}}}
\newcommand{\Tphi}{T_{\phi}}
\newcommand{\Tphim}{\bar{T}_{\phi}}
\newcommand{\tinteract}{t_\mathrm{int}}

\begin{document}
\title{Error Mitigation by Symmetry Verification on a Variational Quantum Eigensolver}
\author{R.~Sagastizabal}
\affiliation{QuTech, Delft University of Technology, P.O. Box 5046, 2600 GA Delft, The Netherlands}
\affiliation{Kavli Institute of Nanoscience, Delft University of Technology, P.O. Box 5046, 2600 GA Delft, The Netherlands}
\author{X.~Bonet-Monroig}
\affiliation{QuTech, Delft University of Technology, P.O. Box 5046, 2600 GA Delft, The Netherlands}
\affiliation{Instituut-Lorentz, Leiden University, The Netherlands}
\author{M.~Singh}
\author{M.A.~Rol}
\author{C.C.~Bultink}
\affiliation{QuTech, Delft University of Technology, P.O. Box 5046, 2600 GA Delft, The Netherlands}
\affiliation{Kavli Institute of Nanoscience, Delft University of Technology, P.O. Box 5046, 2600 GA Delft, The Netherlands}
\author{X.~Fu}
\affiliation{QuTech and Quantum Computer Architecture Lab, Delft University of Technology}
\author{C.H.~Price}
\affiliation{Instituut-Lorentz, Leiden University, The Netherlands}
\author{V.P.~Ostroukh}
\author{N.~Muthusubramanian}
\author{A.~Bruno}
\affiliation{QuTech, Delft University of Technology, P.O. Box 5046, 2600 GA Delft, The Netherlands}
\affiliation{Kavli Institute of Nanoscience, Delft University of Technology, P.O. Box 5046, 2600 GA Delft, The Netherlands}
\author{M.~Beekman}
\affiliation{QuTech, Delft University of Technology, P.O. Box 5046, 2600 GA Delft, The Netherlands}
\affiliation{Kavli Institute of Nanoscience, Delft University of Technology, P.O. Box 5046, 2600 GA Delft, The Netherlands}
\author{N.~Haider}
\affiliation{Netherlands Organisation for Applied Scientific Research (TNO), P.O. Box 155, 2600 AD Delft, The Netherlands}
\affiliation{QuTech, Delft University of Technology, P.O. Box 5046, 2600 GA Delft, The Netherlands}
\author{T.E.~O'Brien}
\affiliation{Instituut-Lorentz, Leiden University, The Netherlands}
\author{L.~DiCarlo}
\affiliation{QuTech, Delft University of Technology, P.O. Box 5046, 2600 GA Delft, The Netherlands}
\affiliation{Kavli Institute of Nanoscience, Delft University of Technology, P.O. Box 5046, 2600 GA Delft, The Netherlands}
\date{\today}

\begin{abstract}
Variational quantum eigensolvers offer a small-scale testbed to demonstrate the performance of error mitigation techniques with low experimental overhead.
We present successful error mitigation by applying the recently proposed symmetry verification technique to the experimental estimation of the ground-state energy and ground state of the hydrogen molecule.
A finely adjustable exchange interaction between two qubits in a circuit QED processor efficiently prepares variational ansatz states in the single-excitation subspace respecting the parity symmetry of the qubit-mapped Hamiltonian.
Symmetry verification improves the energy and state estimates by mitigating the effects of qubit relaxation and residual qubit excitation, which violate the symmetry.
A full-density-matrix simulation matching the experiment dissects the contribution of these mechanisms from other calibrated error sources.
Enforcing positivity of the measured density matrix via scalable convex optimization correlates the energy and state estimate improvements when using symmetry verification, with interesting implications for determining system properties beyond the ground-state energy.
\end{abstract}
\maketitle

Noisy intermediate-scale quantum (NISQ) devices~\cite{Preskill2018NISQ}, despite lacking layers of quantum error correction (QEC), may already be able to demonstrate quantum advantage over classical computers for select problems~\cite{Mca18Review,Cao18Review}.
In particular, the hybrid quantum-classical variational quantum eigensolver (VQE)~\cite{Peruzzo14,Mcc16} may have sufficiently low experimental requirements to allow estimation of ground-state energies of quantum systems that are difficult to simulate purely classically~\cite{Babbush18,Poulin18,Berry18,Kivlichan18}.
To date, VQEs have been used to study small examples of the electronic structure problem, such as $\Htwo$~\cite{OMalley16,Colless18,Ganzhorn18,Hempel18,Kandala17,Kandala18}, HeH+~\cite{Peruzzo14,Shen17}, LiH~\cite{Hempel18,Kandala17,Kandala18}, and BeH$_2$~\cite{Kandala17}, as well as exciton systems~\cite{Santagati18}, strongly correlated magnetic models~\cite{Kandala18}, and the Schwinger model~\cite{Kokail18}. Although these experimental efforts have achieved impressive coherent control of up to 20 qubits, the error in the resulting estimations has remained relatively high due to performance limitations in the NISQ hardware. Consequently, much focus has recently been placed on developing error mitigation techiques that offer order-of-magnitude accuracy improvement without the costly overhead of full QEC.
This may be achieved by using known properties of the target state, e.g., by checking known symmetries in a manner inspired by QEC stabilizer measurements~\cite{Bonet18,Mca18}, or by expanding around the experimentally-obtained state via a linear (or higher-order) response framework~\cite{Mcc17}.
The former, termed symmetry verification (SV), is of particular interest because it is comparatively low-cost in terms of required hardware and additional measurements.
Other mitigation techniques require understanding the underlying error models of the quantum device, allowing for an extrapolation of the calculation to the zero-error limit~\cite{Li17,Temme17,Otten18}, or the summing of multiple calculations to probabilistically cancel errors~\cite{Temme17,Endo18,Huo18}.

In this Rapid Communication, we experimentally demonstrate the use of SV to reduce the error of a VQE estimating the ground-state energy and the ground state of the $\Htwo$ molecule by one order of magnitude on average across the bond-dissociation curve.  Using two qubits in a circuit QED processor, we prepare a variational ansatz state via an exchange gate that finely controls the transfer of population within the single-excitation subspace while respecting the underlying symmetry of the problem (odd two-qubit parity). We show that SV improves the energy and state estimates by mitigating the effect of processes changing total excitation number, specifically qubit relaxation and residual qubit excitation.  We do this through a full density-matrix simulation that matches the experimental energy and state errors with and without SV, and then using this simulation to dissect the contribution of each error source.
Finally, we explore the limitations of SV arising from statistical measurement noise, and find that enforcing the positivity of the fermionic 2-reduced density matrix ties the improvement in energy estimation from SV to the improvement in ground-state fidelity (which was previously not the case).

A VQE algorithm~\cite{Peruzzo14,Mcc16} approximates the ground state $\rhotrue$ of a Hamiltonian $\ham$ by a variational state $\rhoexpt(\vec{\theta})$, with $\vec{\theta}$ a set of parameters that control the operation of a quantum device.
These parameters are tuned by a classical optimization routine to minimize the variational energy $E(\vec{\theta})=\trace[\rhoexpt(\vec{\theta})\ham]$.
In practice, this is calculated by expanding $\rhoexpt(\vec{\theta})$ and $\ham$ over the $N$-qubit Pauli basis $\pauligroup^N:=\{I,X,Y,Z\}^{\otimes N}$,
\begin{equation}
\rhoexpt({\vec{\theta}})=\frac{1}{2^N}\sum_{\hat{P}\in \pauligroup^N }\rhoexpt_{\hat{P}}(\vec{\theta})\hat{P},\hspace{0.5cm}\ham=\sum_{\hat{P} \in \pauligroup^N}h_{\hat{P}}\hat{P}\label{eq:rho_pauli_decomp},
\end{equation}
where the Pauli coefficients are given by $\rhoexpt_{\hat{P}}(\vec{\theta})=\trace[\hat{P}\rhoexpt]$. The variational energy may then be calculated as
\begin{equation}
\Eraw(\vec{\theta})=\sum_{\hat{P} \in \pauligroup^N}\rhoexpt_{\hat{P}}(\vec{\theta})h_{\hat{P}}.
\end{equation}
For example, consider the $\Htwo$ molecule studied in this work.
Mapping the Hamiltonian of this system (in the STO-3G basis) onto four qubits via the Bravyi-Kitaev transformation~\cite{Bravyi02BKmap} and then further reducing dimensions by projecting out two non-interacting qubits~\cite{OMalley16} gives
\begin{align}
\label{eq:ham}
\hamHtwo =& h_{II} II + h_{ZI} ZI + h_{IZ} IZ\nonumber\\ &+ h_{XX} XX + h_{YY} YY + h_{ZZ} ZZ,
\end{align}
where coefficients $h_{\hat{P}}$ depend on the interatomic distance $R$. These coefficients may be determined classically using the OpenFermion~\cite{Openfermion} and psi4~\cite{Parrish17Psi4} packages.
The Pauli coefficients $\rhoexpt_{\hat{P}}$ of the density matrix $\rhoexpt$ are extracted by repeated preparation and (partial) tomographic measurements of the ansatz state.
As one only needs those Pauli coefficients $\rhoexpt_{\hat{P}}$ with non-zero corresponding Hamiltonian coefficients $h_{\hat{P}}$, one need not perform full tomography of $\rhoexpt$.
However, in a small-scale experiment, full state tomography of $\rhoexpt$ may still be feasible, and may provide useful information for the purposes of benchmarking.
In particular, the fidelity of $\rhoexpt$ to $\rhotrue$,
\begin{equation}
\label{eq:statefidel}
\Fidel^{\mathrm{(raw)}}= \trace[\rhoexpt \rhotrue],
\end{equation}
is a  more rigorous measure of the ability to prepare the ground state than the energy error,
\begin{equation}
\label{eq:enegyerror}
\Eerr^{\mathrm{(raw)}} = \trace\left[\left(\rhoexpt-\rhotrue\right) \ham\right].
\end{equation}
Error mechanisms such as decoherence pull $\rhoexpt$ away from $\rhotrue$, decreasing $\Fidel$ and increasing $\Eerr$.

These errors may be mitigated by using internal symmetries $\hat{S}\in\pauligroup^N$~\footnote{As described in Refs.~\cite{Bonet18,Mca18}, one does not require $\hat{S}$ to be a Pauli operator, however this makes the SV procedure significantly simpler.} of the target problem, such as parity checks~\cite{Mca18,Bonet18}.
These checks project $\rhoexpt$ to a symmetry verified matrix $\rhosv$ that lies in the $\langle\hat{S}\rangle=s$ subspace of the symmetry.
This projection could be performed via direct measurement of $\hat{S}$ on the quantum device, but one may instead extract the relevant terms of the density matrix $\rhosv$ in post-processing:
\begin{equation}
\rhosv_{\hat{P}} = \frac{\rhoexpt_{\hat{P}}+s\rhoexpt_{\hat{S}\hat{P}}}{1+s\rhoexpt_{\hat{S}}}\label{eq:SQSE},
\end{equation}
The right-hand side may be obtained by partial tomographic measurement of the ansatz state, with at most twice the number of Pauli coefficients that need to be measured.
This upper bound is not always achieved.
For example, the $\hamHtwo$ Hamiltonian has a $\hat{S}=ZZ$ symmetry, which maps the non-zero Pauli terms in $\hamHtwo$ to other non-zero Pauli terms in $\hamHtwo$.
Symmetry verification in this problem then does not require any additional measurements to estimate $\Esv$ beyond those already required to estimate $\Eraw$.
Even when it does require additional measurements, SV remains attractive because it does not require additional quantum hardware or knowledge of the underlying error model.
One can show that the SV state $\rhosv$ may be equivalently obtained via a variant of the quantum subspace expansion (QSE)~\cite{Mcc17}, suggesting an alternative name of S-QSE~\cite{Bonet18}.

One may further minimize the error in a quantum algorithm by tailoring the quantum circuit or the gates within.
In a VQE, one wishes to choose a variational ansatz motivated by the problem itself~\cite{OMalley16,McCleanplat18} while minimizing the required quantum hardware~\cite{Kandala17}.
To balance these considerations, we suggest constructing an ansatz from an initial gate-set that is relevant to the problem at hand.
For example, in the electronic structure problem, the quantum state is generally an eigenstate of the fermion number.
When mapped onto qubits, this often corresponds to a conservation of the total qubit excitation number.
Gates such as single-qubit $Z$ rotations, two-qubit C-Phase~\cite{DiCarlo09}, and two-qubit $\iSWAP$~\cite{Majer07} gates preserve this number, making these gates a good universal gate set (within the target subspace~\cite{Brod14}) for quantum simulation of electronic structure.
In the example of $\Htwo$, the total two-qubit parity ($ZZ$) is indeed conserved and the ground state at any $R$ may be generated by applying to $\ket{01}$ or $\ket{10}$ an exchange gate
\begin{equation}
\iSWAPtheta = \begin{pmatrix}
1 & 0 & 0 & 0\\
0 & \cos\theta & i\sin\theta & 0\\
0 & i\sin\theta & \cos\theta & 0\\
0 & 0 & 0 & 1\\
\end{pmatrix}
\label{eq:iswap_unitary}
\end{equation}
with $R$-dependent optimal exchange angle $\theta$ and a follow-up phase correction on one qubit.

\begin{figure}
  \begin{center}
    \includegraphics[width=0.5\textwidth,trim={0 0.85cm 0 1cm},clip]{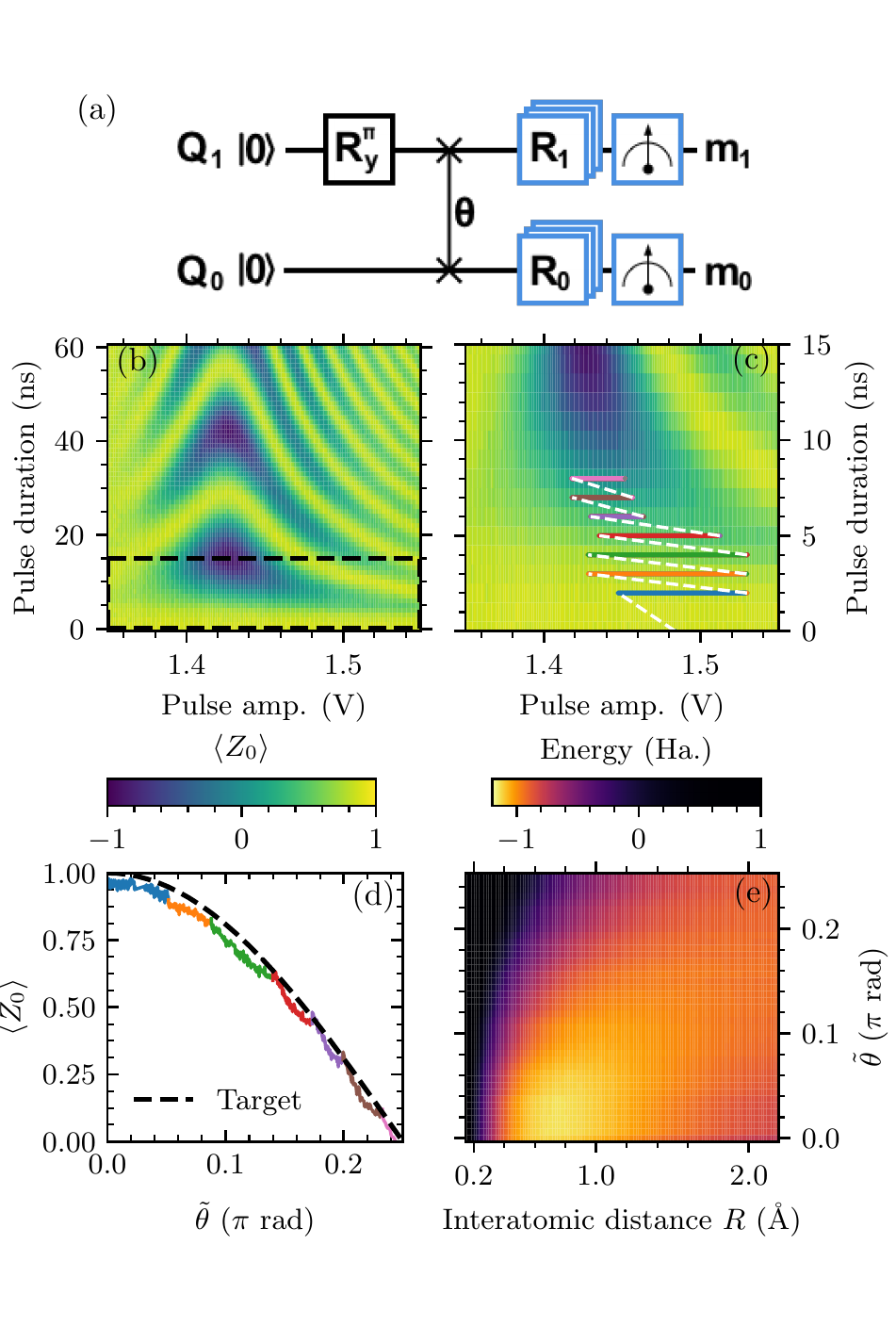}
  \end{center}
  \caption{\label{fig:ISwap_gate}Quantum circuit and energy landscape of the variational eigensolver.
(a) Quantum circuit for generating and measuring the variational ansatz state.
(b) Coherent excitation exchange, produced as $\Qzero$ is fluxed into resonance with $\Qone$ by a square flux pulse. Pulse amplitude ($x$ axis) parametrizes the frequency to which $\Qzero$ is flux pulsed ($\sim1.428~\V$ bringing it on resonance with $\Qone$).
(c) Zoom-in of (b) into the region used in the experiment to control the exchange of population between $\Qzero$ and $\Qone$. Colored lines illustrate the hybrid path in pulse duration and amplitude that maps out a finely-adjustable $\tilde{\theta}$ range.
(d) Excitation of $\Qzero$ along the hybrid path, showing the matching of the experimentally-defined $\tilde{\theta}$ to the target $\theta$ defined in Eq.~(\ref{eq:iswap_unitary}) (black dashed curve). Colors [matching (c)] illustrate different pulse durations used in each segment.
(e) Landscape of energies $\Eraw(\tilde{\theta},R)$ as function of the experimentally-defined $\tilde{\theta}$ angle and the interatomic distance $R$.
}
\end{figure}

We now experimentally investigate the benefits of SV in the VQE of $\Htwo$ using two of three transmon qubits in a circuit QED quantum processor (see details in~\cite{SOM_VQE}).
The two qubits ($\Qzero$ and $\Qone$) are coupled by a common bus resonator, and have dedicated microwave lines for single-qubit gating, flux bias lines for local and ns-scale control of their frequency, and dedicated readout resonators coupling to a common feedline for independent readout by frequency multiplexing.
We prepare the ansatz state with an efficient circuit [Fig.~\ref{fig:ISwap_gate}(a)] that first excites $\Qone$ with a $\pi$ pulse to produce the state $\ket{10}$, and then flux pulses $\Qzero$ into resonance with $\Qone$ to coherently exchange the excitation population.
A sweep of flux-pulse amplitude and duration [Fig.~\ref{fig:ISwap_gate}(b)] reveals the expected chevron pattern that is the hallmark of coherent population exchange between the two qubits, albeit with some asymmetry arising from imperfect compensation of linear distortion in the flux-bias line.
To finely control population exchange without being limited by the $1~\ns$ resolution in pulse duration, we stitch together a hybrid path in pulse duration and amplitude.
This results in a fine experimental knob $\tilde{\theta}$ (1500 possible settings) that controls population exchange like $\theta$ in Eq.~(\ref{eq:iswap_unitary}) [Fig.~\ref{fig:ISwap_gate}(c)], although with additional single-qubit phases.
The circuit concludes with simultaneous pre-rotation gates on both qubits followed by simultaneous measurement of both qubits, in order to perform tomography of the prepared ansatz state.
To fully reconstruct the state, we use an overcomplete set of 36 pre-rotation pairs and extract estimates of the average measurement for each qubit as well as their shot-to-shot correlation using $\Nmeas$ measurements per pre-rotation.
Note that single-qubit phase corrections are not required immediately following the exchange gate, as phase rotations can be performed virtually from the fully-reconstructed state.

\begin{figure}
  \begin{center}
    \includegraphics[width=\columnwidth]{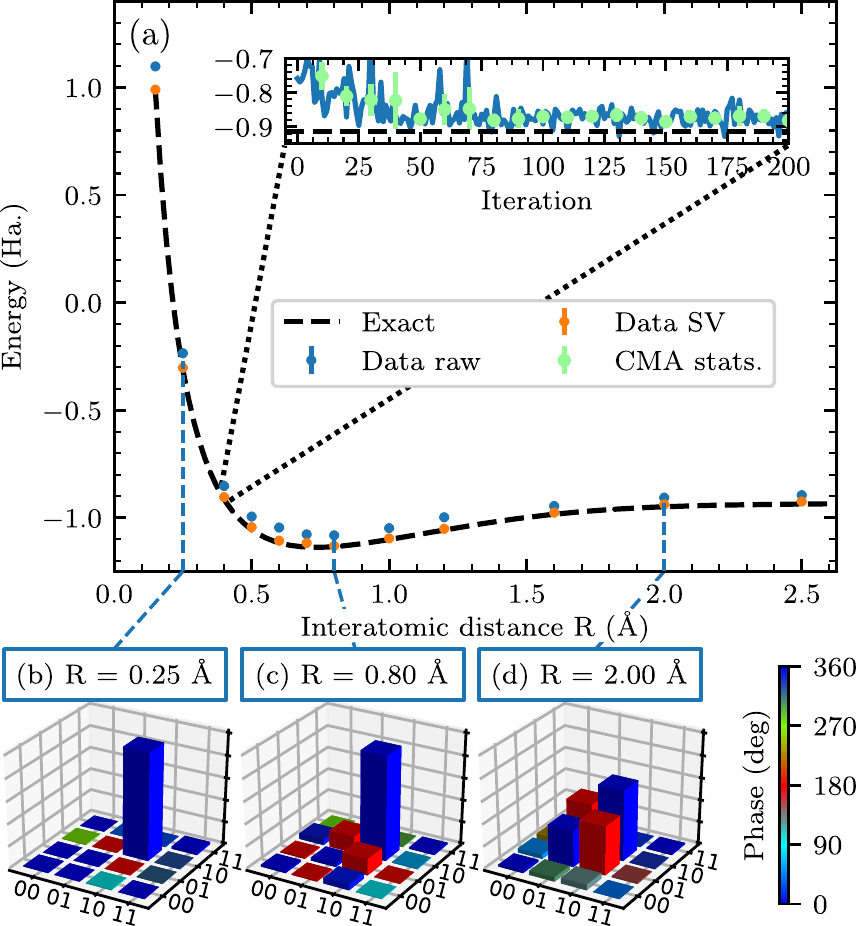}
  \end{center}
  \caption{\label{fig:VQE_convergence}Convergence of the VQE algorithm.
  (a) Experimental VQE estimate of $\Htwo$ ground-state energy as a function of interatomic distance $R$.
  At each chosen $R$, we minimize the raw energy $\Eraw$ (blue data points) over the variational parameter $\tilde{\theta}$ using the CMA-ES evolutionary algorithm~\cite{Hansen09}.
  Applying SV to the converged solution (orange data points) lowers the energy estimate towards the exact solution (dashed curve).
  Inset: A typical optimization trace for the convergence of the energy estimate.
  (b-d) The reconstructed density matrices of the converged states at (b) $R=0.25~$\r{A}, (c) $R=0.80~$\r{A}, and (d) $R=2.00~$\r{A}, showing that the converged states lie mostly in the single-excitation subspace, and that entanglement increases with the interatomic distance $R$.}
\end{figure}

We now optimize the VQE to approximate the ground-state energy and ground state of $\Htwo$.
At each chosen $R$, we employ the covariance matrix adaptation evolution strategy (CMA-ES) optimization algorithm~\cite{Hansen09}, using $\Eraw$ as cost function and $\tilde{\theta}$ as single variational parameter.
The evolutionary strategy optimizes $\tilde{\theta}$ over repeated generations of $N_{\mathrm{pop}}=10$ samples of $\Eraw(\tilde{\theta})$, each calculated from a raw density matrix $\rhoexpt$ using linear inversion of $\Nmeas=10^3$.
A typical optimization [Fig.~\ref{fig:VQE_convergence}(a) inset] converges after $\sim 20$ generations ($\sim 2~\mathrm{hours}$). The converged state is finally reconstructed with greater precision, using $\Nmeas=10^5$. Figure~2 shows the resulting energy estimate for twelve values of $R$ and the reconstructed optimized state at three such distances. These tomographs show that the optimal solutions are concentrated in the single-excitation subspace of the two qubits, with two-qubit entanglement increasing as a function of $R$.

\begin{figure}
  \begin{center}
    \includegraphics[width=\columnwidth]{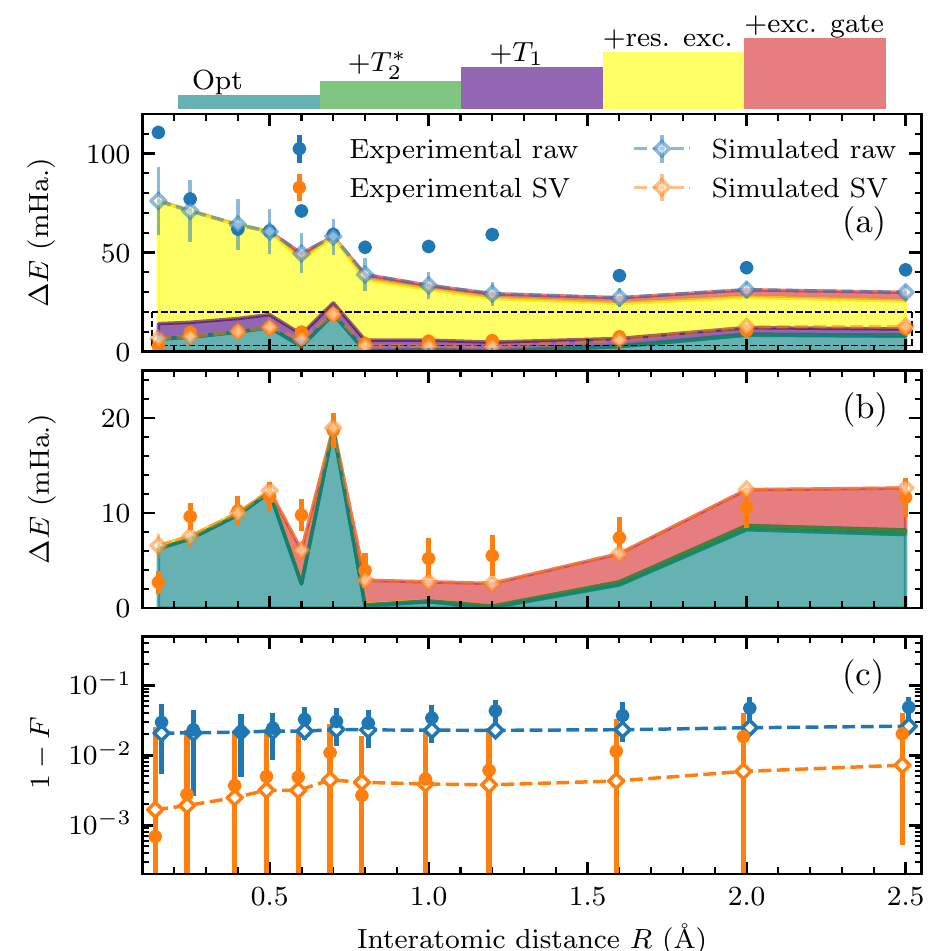}
  \end{center}
  \caption{Impact of SV in ground-state energy and state fidelity, and dissected error budget.
  (a) Experimental (solid circles) energy error $\Eerr$ without and with SV compared to the result (empty circles) of a full density-matrix simulation using the full error model. The contributions from optimizer inaccuracy, qubit dephasing, qubit relaxation, residual qubit excitations
  and increased $\Qzero$ dephasing during the exchange gate are shown as shaded regions for the case of no SV applied. Without SV, $\Eerr$ is clearly dominated by residual qubit excitation.
  (b) Zoom-in on experimental and simulated $\Eerr$ with SV and corresponding error budget. With SV, the effects of residual excitation and qubit relaxation are successfully mitigated, as predicted in Ref.~\onlinecite{Bonet18}. The remaining energy error is dominated
  by optimizer inaccuracy. Simulation error bars are obtained by modelling measured fluctuations of $\Tone$, $\Ttwos$, and residual excitation.
  (c) Experimental (solid circles) infidelity to the true ground state without and with SV compared to simulation using the full error model (empty circles).}
  \label{fig:SV_impact_fig}
\end{figure}

Performing the described symmetry verification procedure on the converged states shows improvement across the entire bond-dissociation curve.
To quantify the improvement, we focus on the energy error $\Eerr$ and the infidelity $1-\Fidel$ to the true ground state, with and without SV (Fig.~\ref{fig:SV_impact_fig}).
SV reduces the energy error by an average factor $\sim10$ and reduces the infidelity by an average factor $\sim9$.
In order to quantitatively understand the limits of the VQE optimization, and to clearly pinpoint the origin of the SV improvement, we simulate the experiment via the density-matrix simulator \emph{quantumsim}~\cite{Obrien17}, using an error model built from independently measured experimental parameters~\cite{SOM_VQE}.
We build the error model incrementally, progressively adding: optimization inaccuracy (the difference between the state ideally produced by the converged $\theta$ and the true ground state);
dephasing on both qubits (quantified by the measured Ramsey dephasing times $\Ttwos$); relaxation on both qubits (quantified by the measured relaxation times $\Tone$); residual qubit excitations (measured from single-shot histograms with each qubit prepared in $\ket{0}$); and increased dephasing of $\Qzero$ during the exchange gate (quantified by its reduced $\Ttwos$ when tuned into the exchange interaction zone).
By plotting the errors from each increment of the model, we are able to dissect the observed experimental error into its separate components without [Fig.~\ref{fig:SV_impact_fig}(c)] and with [Fig.~\ref{fig:SV_impact_fig}(b)] SV.
Measured temporal fluctuations of dephasing, relaxation and residual excitation are used to obtain simulation error bars.

The simulation using the full error model shows fairly good matching with experiment for both the ground-state energy error [Figs.~\ref{fig:SV_impact_fig}(a,b)] and the state infidelity [Fig.~\ref{fig:SV_impact_fig}(c)], with and without SV.
The error model dissection shows that the energy error improvement from SV results from the mitigation of errors arising from qubit relaxation and residual qubit excitations. This is precisely as expected: these error mechanisms change total qubit excitation number and violate the underlying $ZZ$ symmetry. Using SV changes the dominant error mechanism from residual qubit excitation to optimization inaccuracy. This error could be reduced experimentally by increasing $\Nmeas$ during the optimization, at the cost of increased convergence time. The improvement in state infidelity by SV can be explained along similar lines.
We observe some increased deviations between the observed and simulated state infidelity at large $R$. We attribute these to limitations in our to modeling of error during the exchange gate (whose duration increases with $R$).

\begin{figure}[h]
  \begin{center}
    \includegraphics[width=\columnwidth]{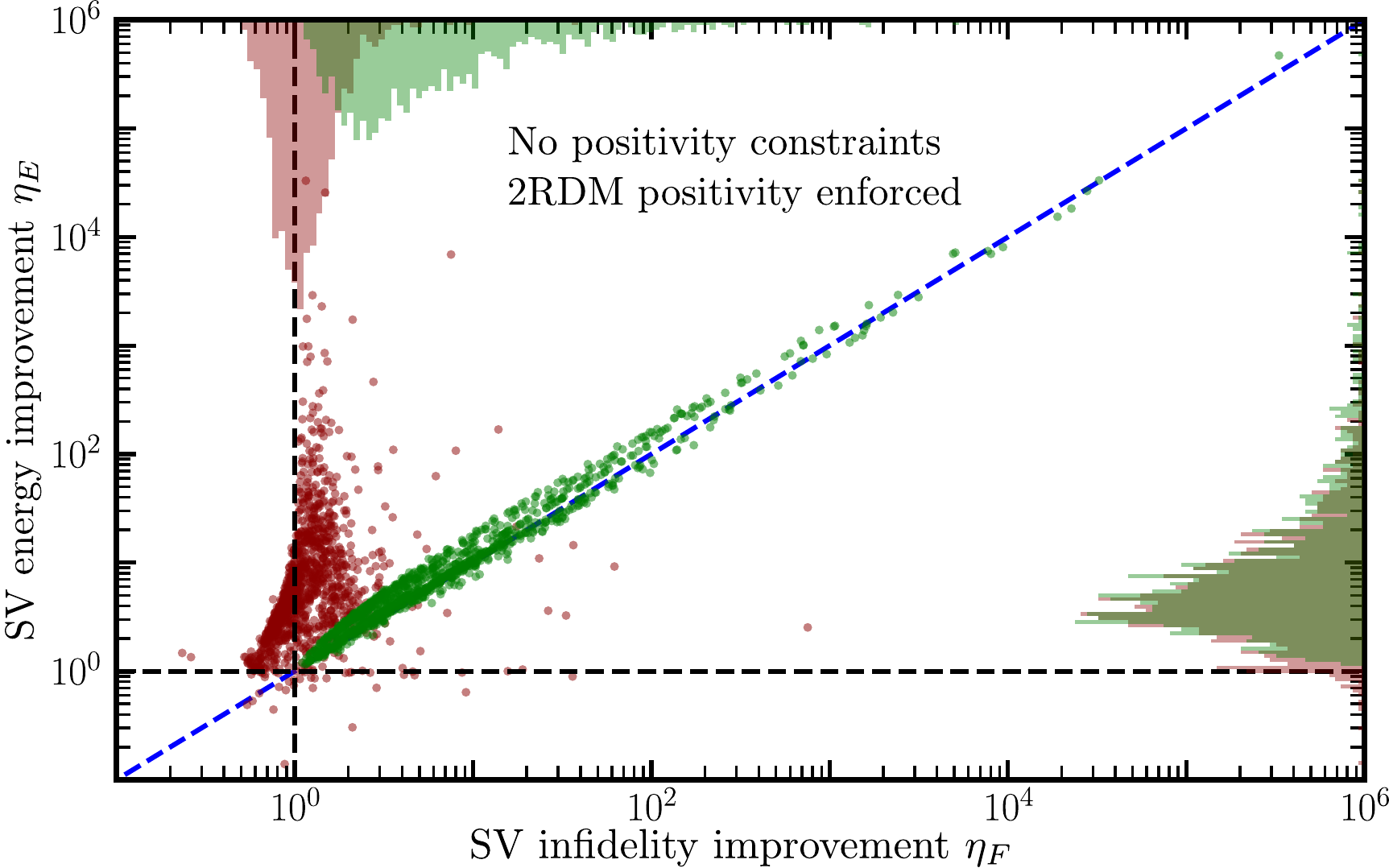}
  \end{center}
  \caption{Constraining positivity with symmetry verification to mitigate the effect of sampling noise.
  The experimental data from Fig.~\ref{fig:SV_impact_fig} is split into $100$ sample simulations for each $R$, increasing the sampling noise by a factor of $10$ and making it comparable to other sources of experimental error.
  For each sample, we plot (red) the relative energy error and infidelity [Eq.~(\ref{eq:relative})].
  Values below $1$ (dashed lines) indicate that SV has not provided an improvement, as may be the case when the density matrix has negative eigenvalues.
  We restore the improvement from SV by constraining the positivity of the 2-reduced density matrix~\cite{Rubin18} (green).
  Histograms on the top and right axes show the marginal distribution of the two scatter plots.
  When the density matrices are constrained to be positive, we observe the points fall along the line $y=x$ (blue dashed line), indicating that SV improves both metrics by the same amount.
}
  \label{fig:physicality}
\end{figure}

VQEs rely on variational bounding to ensure that the obtained approximation to the ground-state energy is accurate, but this is only guaranteed when the experimental results correspond to a physical state.
Our method for calculating the ground-state energy [Eq.~(\ref{eq:rho_pauli_decomp})] independently estimates each Pauli coefficient of the density matrix with error $\propto \Nmeas^{-1/2}$.
Such estimation cannot guarantee a set of Pauli coefficients that could have come from a positive density matrix.
This in turn breaks the variational lower bound on the energy estimate, and increases the error in estimates of other properties of the true ground state~\cite{Rubin18,Blume06}.
As experimental error is reduced, $\rhoexpt$ tends towards a rank-$1$ density matrix, increasing its chance of being unphysical~\cite{Blume06}.
Moreover, $\rhosv$ is a lower-rank density matrix than $\rhoexpt$ (being projected onto a subspace of the Hilbert space), which implies that unphysicality may be enhanced by SV.
The variance in a given term $\rho_{\hat{P}}$ post-SV can be calculated as
\begin{equation}
\mathrm{Var}[\rhosv_{\hat{P}}]\approx \frac{3\Nmeas}{\Nmeas(1+\trace[\rhoexpt\hat{S}])}.
\end{equation}
SV has maximal impact on the quantum state precisely when this denominator is small, so this represents a natural bound for the power of SV as an error mitigation strategy.

The effect of sampling noise may be mitigated somewhat by restricting the fermionic 2-reduced density matrix to be positive (which may be completed in polynomial time)~\cite{Rubin18}.
To investigate the effect of such mitigation, we bin the data used for final tomography of converged states to construct 100 density matrices with $\Nmeas=10^3$ at each $R$, thus increasing the sampling noise by a factor of $10$.
We wish to study the relative improvement of SV in the two figures of merit, which we quantify as
\begin{equation}
\eta_{\mathrm{E}}=\frac{|\Delta E^\textrm{(raw)}|}{|\Delta E^\textrm{(SV)}|}\ \mathrm{and}\ \eta_{\mathrm{F}}=\frac{|1-F^{(\textrm{raw})}|}{|1-F^{(\textrm{SV})}|},
\label{eq:relative}
\end{equation}
when physicality of the raw density matrices is enforced and not.
To enforce physicality, we employ a convex optimization routine to find the closest positive semidefinite matrix to the experimentally measured $\rhoexpt$ (closest in the $L^2$ norm sense on the space induced by the the Pauli basis).
We then apply symmetry verification to the post-processed density matrix.
Figure~\ref{fig:physicality} shows a scatter plot of $\eta_{\mathrm{E}}$ and $\eta_{\mathrm{F}}$, and relative histograms of each.
Without enforcing physicality, SV makes no significant improvement to the state fidelity, although it almost always improves the energy error.
However, when positivity is enforced, SV greatly improves the overlap with the true ground state.
We also find that the improvement in the energy from SV is equal to the improvement in fidelity when the starting state is physical, but is relatively uncorrelated when the starting state is not.
This makes sense, as the energy gain from SV given a physical matrix comes directly from substituting higher energy states with density on the ground state.
It is unclear whether such a strong trend will continue in larger systems without requiring too stringent a positivity constraint.
As this is a four-orbital two-electron system, enforcing the positivity of the 2-reduced density matrix enforces positivity on the entire density matrix (which is exponentially difficult in the system size~\cite{Liu07}).
Testing this scalability is a clear direction for future research~\footnote{Note that, for this system, enforcing positivity of the $1$-reduced density matrix corresponds to ensuring that all expectation values are bounded between $-1$ and $1$, and so this does not provide any additional data.}.

In summary, we have experimentally demonstrated the use of SV to mitigate errors in the VQE of $\Htwo$ with two transmon qubits.
We implemented an efficient variational ansatz based on an exchange gate producing finely adjustable population transfer in the single-excitation subspace, respecting the $ZZ$ symmetry of the $\Htwo$ Hamiltonian. Verification of this symmetry reduced the error of the estimated ground-state energy and the ground state by one order of magnitude on average over the full dissociation curve. A full density-matrix simulation of our system allowed us to budget the contributions from known experimental error mechanisms.
We observe that SV mitigates the effect of processes that affect total qubit excitation number, specifically qubit relaxation and residual excitation.
Finally, we have investigated the effect of reconstructing density matrices via linear tomographic inversion in the presence of sampling, which voids the guarantee of positivity and in turn the guarantee that SV improves estimation of the ground state.
Intriguingly, we observe that when physicality is enforced, the reduction in energy error from SV is directly linked to the increase in fidelity to the ground state.
If this observation extends to larger systems, a user can be confident that symmetry-verified Pauli coefficients are accurate for calculations beyond the ground-state energy.

\begin{acknowledgments}
We acknowledge experimental assistance from F.~Luthi, B.~Tarasinski and C.~Dickel, theoretical assistance from B.~Terhal and V.~Cheianov, assistance with psi4 from L.~Visscher and B.~Senjean, and useful discussions with B.~Varbanov, F.~Malinowski, Y.~Herasymenko, M.~Steudtner, and C.W.J.~Beenakker. This research is funded by an ERC Synergy Grant, the Netherlands Organization for Scientific Research (NWO/OCW), the China Scholarship Council, and IARPA (U.S. Army Research Office grant W911NF-16-1-0071). T.E.O'B. is additionally funded by Shell Global Solutions BV.
\end{acknowledgments}

\renewcommand{\theequation}{S\arabic{equation}}
\renewcommand{\thefigure}{S\arabic{figure}}
\renewcommand{\thetable}{S\arabic{table}}
\setcounter{figure}{0}
\setcounter{equation}{0}
\setcounter{table}{0}
\section*{Supplemental material for "Error mitigation by Symmetry Verification on a Variational Quantum Eigensolver"}
\date{\today}
\maketitle
\section{Device fabrication}
A high-resistivity intrinsic silicon wafer was cleaned with acetone and 2-isopropanol, and stripped of native oxides using buffered oxide etch solution (BOE $7:1$). The wafer was subjected to HMDS vapor and sputtered with $200~\nm$ of NbTiN followed by dicing into smaller dies. The device plane was spun with a high-contrast positive tone resist and patterned using e-beam lithography. The exposed base layer was subtractively patterned using reactive ion etching and the resist was stripped. This was followed by spinning of a bilayer resist for fabrication of Josephson junctions by double-angle shadow evaporation. For the fabrication of airbridges, a $6~\um$ thick e-beam resist was patterned and subjected to reflow. A $450~\nm$ thick layer of aluminum was deposited using an e-beam evaporator. The chip was diced and wirebonded to a printed circuit board.

\section{Experimental setup}
The device was mounted on a copper sample holder attached to the mixing chamber of a Leiden Cryogenics CF-650 dilution refrigerator with $\sim22~\mK$ base temperature. For radiation shielding, the cold finger was enclosed by a copper can coated with a mixture of Stycast 2850 and silicon carbide granules ($15$ to $1000~\nm$ diameter) used for infrared absorption. To shield against external magnetic fields, the can was enclosed by an aluminum can and two Cryophy cans. Microwave lines were filtered using $60~\dB$ of attenuation with both commercial cryogenic attenuators and home-made Eccosorb filters for infrared absorption. Flux-bias lines were also filtered using commercial low-pass filters and Eccosorb filters with a stronger absorption. Fast flux-pulses were coupled to the flux-bias lines via room-temperature bias tees.

Amplification of the readout signal was done in three stages: first a TWPA (provided by MIT-LL~\cite{Macklin307}) located at the mixing chamber plate, then a Low Noise Factory HEMT at the $4~\K$ plate, and finally a Miteq amplifier at room temperature.
The TWPA was mounted on a separate sample holder with the same shielding layers as the device.

Room-temperature electronics used both commercial hardware and custom hardware developed in QuTech.
Rhode \& Schwarz SGS100 sources provided all microwave signals for single-qubit gates and readout. The DC bias was provided by home-built current sources (IVVI racks). QuTech arbitrary waveform generators (QWG) generated the modulation envelopes for single-qubit gates and the flux pulse for the exchange gate. A Zurich Instruments UHFQA was used to perform independent readout of both qubits as well as their correlation. QuTech mixers were used for frequency up- and down-conversion. The QuTech Central Controller Light (CCL) coordinated the triggering of QWGs and UHFQA.

\begin{figure*}[h]
 \begin{center}
   \includegraphics[width=0.8\textwidth]{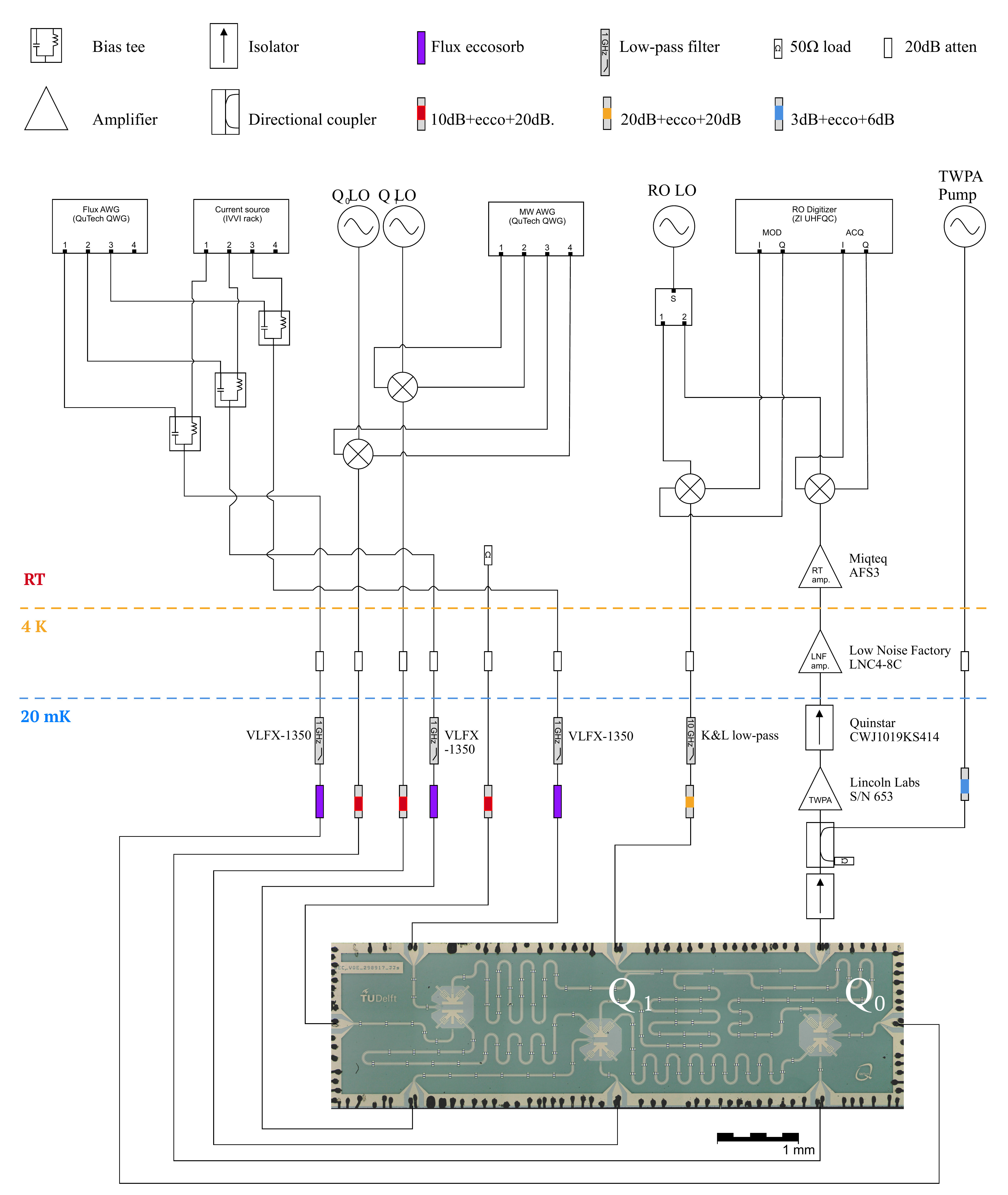}
 \end{center}
 \caption{Device and wiring schematic. The $2~\mm \times 7~\mm$ chip contains three Starmon~\cite{Versluis17} qubits. Qubit pairs $\Qzero$-$\Qone$ and $\Qone$-$\Qtwo$ are coupled by bus resonators. Qubits have dedicated microwave drive lines, flux bias lines, and dispersively-coupled readout resonators. Readout resonators are coupled to a common feedline allowing independent qubit readout by frequency multiplexing. See text for details of cryogenic system and wiring. In this experiment, we only make use of $\Qzero$ and $\Qone$. The unused leftmost qubit, $\Qtwo$, is parked at its sweetspot frequency ($4.128~\GHz$) throughout.}
\end{figure*}

All measurements were controlled at the software level with qCoDeS~\cite{QCoDeS16} and PycQED~\cite{PycQED16} packages. The QuTech OpenQL compiler translated high-level Python code into the eQASM code~\cite{Fu19} forming the input to the CCL.

\section{Measured device parameters}
\begin{table}[h]
\begin{tabular}{ l | c | c  }
\hline
Qubit & $Q_1$ & $Q_0$\\
\hline
Readout resonator frequency (GHz) & $8.0005$ & $7.7377$ \\
Qubit sweetspot frequency (GHz) & $5.1468$ & $5.9207$\\
$\Tone$ $\left(\mu\textrm{s}\right)$ & $9.8\pm1.0$ & $11.7\pm0.6$\\
$\Ttwos$ $\left(\mu\textrm{s}\right)$ & $9.0\pm1.3$ & $17.3\pm1.0$\\
Residual qubit excitation $(\%)$ & $1.34\pm 0.20$ & $0.25\pm 0.09$\\
Single-qubit gate fidelity & $98.6\%$ & $99.1 \%$\\
Coupling $\frac{J_1}{2\pi}$ (MHz) & \multicolumn{2}{c}{$20.9$}\\
\hline
\end{tabular}
\caption{Measured device parameters. Resonator and qubit frequencies were measured by spectroscopy, while relaxation and dephasing times, $\Tone$ and $\Ttwos$, respectively, were measured by standard time-domain experiments.  Error bars on $\Tone$ and $\Ttwos$ correspond to the standard deviation of 56 repeated measurements performed over a 24-hour period. See text for the procedure used to quantify residual qubit excitations. Single-qubit gate fidelity was measured by randomized benchmarking. The qubit-qubit coupling strength was measured both by spectroscopy and time-domain measurements.}
  \label{table:specs}
\end{table}

\begin{figure}[t!]
  \begin{center}
    \includegraphics[width=\columnwidth]{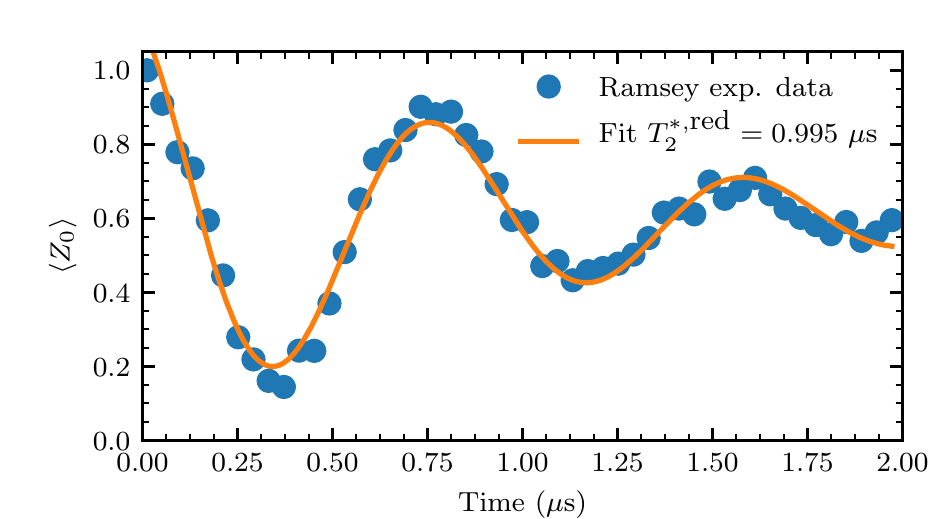}
  \end{center}
  \caption{Ramsey experiment with qubit $\Qzero$ DC-flux biased to the frequency of the resonant exchange interaction ($5.1468~\GHz$, the $\Qone$ sweetspot frequency), but with qubit $\Qone$ biased away to a lower frequency. The best fit value $\Ttwosred = 0.995~\us$ is used for modelling the increased dephasing of $\Qzero$ during the exchange gate.}
  \label{fig:SOM_Ttwosred}
  \end{figure}

We ran a series of characterization experiments to extract the device parameters needed as inputs to the error model used in our density-matrix simulaiton. These are summarized in Table~\ref{table:specs}. The qubit relaxation time $\Tone$ and dephasing time  $\Ttwos$ for each qubit were measured using standard time-domain sequences. The reduced dephasing time $\Ttwosred$ of $\Qzero$ during the exchange gate was measured by DC biasing $\Qzero$ to $5.1468~\GHz$ while DC biasing $\Qone$ sufficiently far away from its sweetspot. We extract  $\Ttwosred=0.995~\us$ from a standard Ramsey time-domain experiment (Fig.~\ref{fig:SOM_Ttwosred}). Single-qubit gate fidelity was extracted from randomized benchmarking of each qubit separately.

\begin{figure*}[b!]
  \begin{center}
    \includegraphics[width=\textwidth]{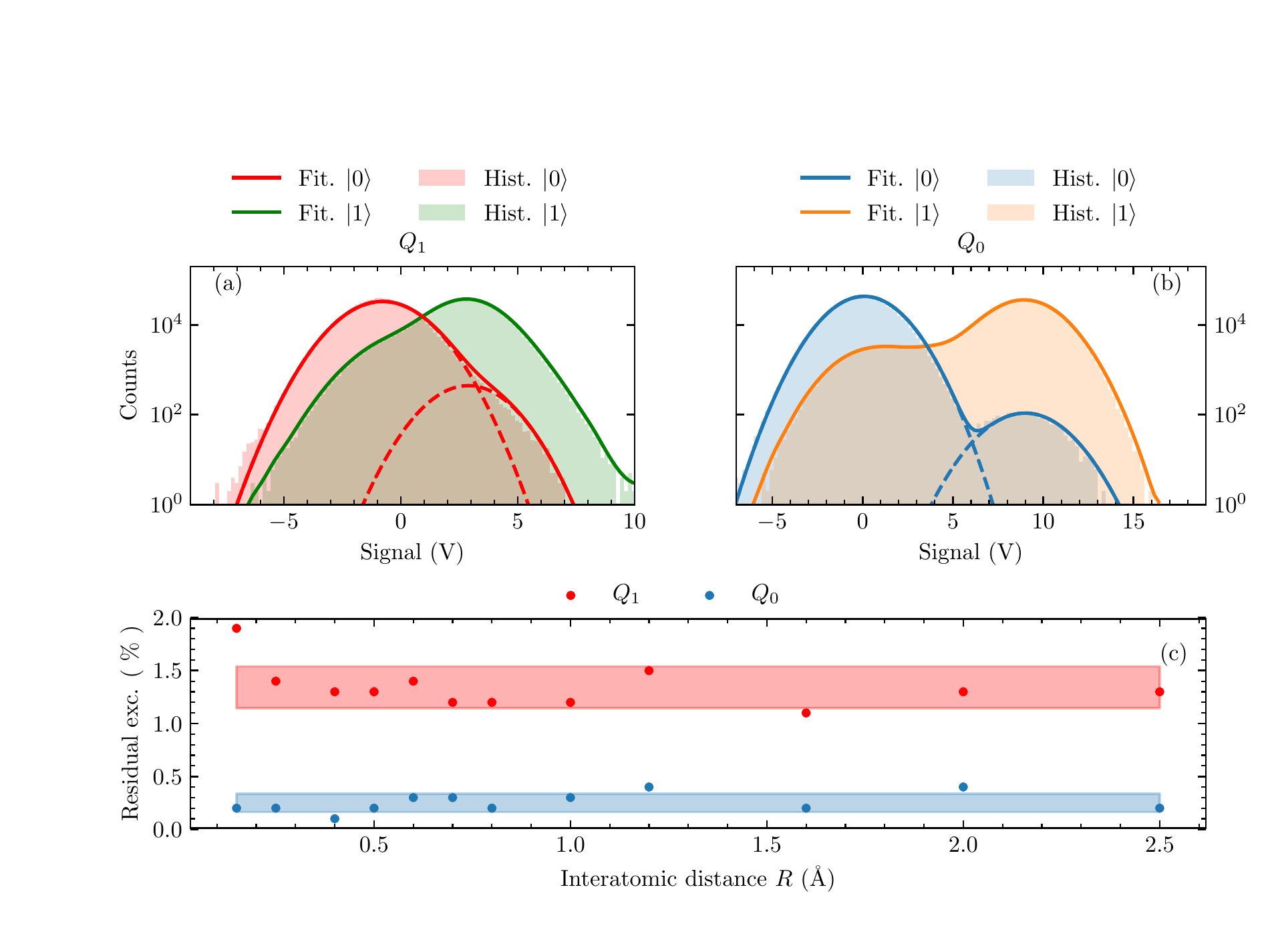}
  \end{center}
  \caption{Residual qubit excitations. Single-shot readout histograms for (a) $\Qone$ and (b) $\Qzero$ nominally prepared in $\ket{0}$ and $\ket{1}$. We extract a residual qubit excitation from the best-fit double gaussian on the ground-state histogram (solid curves). (c) Residual excitations extracted from calibration data in the final, post-VQE-convergence in the dataset of Figs.~2 and 3 at each value of $R$. The average (standard deviation) of the residual excitation is $0.25\%$ ($0.09\%$) for $\Qzero$ and $1.3\%$ ($0.2\%$) for $\Qone$.}
  \label{fig:SOM_res}
\end{figure*}

We quantified residual qubit excitations from a subset of the measurement set used to calibrate the measurement operators in the post-convergence tomographic reconstruction at each value of $R$ in Fig.~\ref{fig:SV_impact_fig} (see section below). The measurement set consists of $7\times10^5$ measurements with the two qubits nominally prepared in each of the four computational states $\ket{00}$, $\ket{01}$, $\ket{10}$, and $\ket{11}$.  Since all homodyne voltage shots were stored (not just their average), we could construct histograms of the measurements for $\ket{00}$  before 1-bit digitization. The residual excitation was then extracted from a double Gaussian fit  [Fig.~\ref{fig:SOM_res}(a-b)]. We performed this procedure to extract residual qubit excitations for  every data point in Fig.~\ref{fig:SOM_res}(c).

\section{Tomographic reconstruction and limitations}
Tomographic reconstruction was performed with the same technique described in~\cite{Saira14}.
We provide a brief description here for completeness.
For each measurement channel (measurement of $\Qone$, measurement of $\Qzero$, and their correlation), the average measurement outcome is given by $\langle m_i \rangle=\trace(\hat{M_i}\rho )$, with operator

\begin{equation}
\hat{M_i} = \beta_{II}^{i} \hat{II} + \beta_{IZ}^{i} \hat{IZ} + \beta_{ZI}^{i} \hat{ZI} + \beta_{ZZ}^{i} \hat{ZZ},
\label{eq:meas_tomo}
\end{equation}
and real-valued coefficients  $\beta_{j}^{i}$.
Single-shot measurements of $\Qzero$ and $\Qone$ are 1-bit digitized before correlation and before averaging each of the three channels.

The simultaneously applied measurement pre-rotations $\Rzero$ and $\Rone$ consist of the 36 pairs created by drawing each rotation separately from the set $\{ I, X_{\pi},X_{\pi/2},Y_{\pi/2},X_{-\pi/2},Y_{-\pi/2}\}$. These measurement pre-rotations effectively change the measurement operator to
\[
M_{i}^{k,l} = \trace \left( R^{k,l,\dagger } \hat{M_{i}} R^{k,l} \right).
\]
There are thus 108 linear equations (36 per channel) linking the averaged measurement to the 15 nontrivial 2-qubit Pauli coefficients (we force $\langle\hat{II}\rangle=1$). We then extract the Pauli coefficients by performing least-squares linear inversion.
Prior to the linear inversion, the measurements are scaled to approximately match the noise in the three channels.

The coefficients $\beta_{j}^{i}$ are obtained from standard calibration measurements. The two qubits are nominally prepared in the four computational states and measured. In total, we perform $7\times \Nmeas$ measurements per computational state. The matrix relating the four measurement averages of a channel to the coefficients has elements of the form $\langle\hat{II}\rangle$, $\pm \langle\hat{IZ}\rangle$, $\pm \langle\hat{ZI}\rangle$ and $\pm \langle\hat{ZZ}\rangle$. By taking into account the calibrated residual qubit excitations, which reduce the magnitude of $\langle\hat{IZ}\rangle$, $\langle\hat{ZI}\rangle$, and $\langle\hat{ZZ}\rangle$ from unity, we ensure that the coefficients $\beta_{j}^{i}$ and thus also the operator $M_i$ are not corrupted by residual excitation.

\begin{figure}[t!]
  \begin{center}
    \includegraphics[width=\columnwidth]{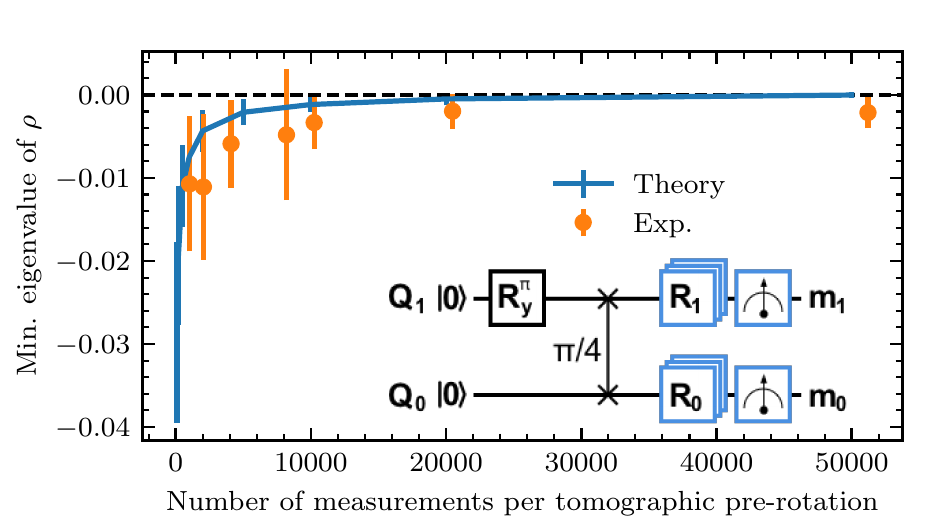}
  \end{center}
  \caption{Minimal eigenvalue of density matrices obtained from linear tomographic reconstruction with different $\Nmeas$. Here, the state preparation targets a Bell state using our variational ansatz (inset) with $\theta =\pi/4$ (producing a $\sqrt{\textrm{iSWAP}}$ gate). Physicality constraints on density matrices restrict their eigenvalues to be non-negative. We observe negative minimum eigenvalues over the entire range of $\Nmeas$. A \emph{quantumsim} simulation produces a similar trend, asymptoting to a  physical state by $\Nmeas\sim 50,000$.
 }
  \label{fig:SOM_negativity}
\end{figure}

Tomography by linear inversion does not ensure physicality of the reconstructed density matrix. We investigate this effect
by performing tomography with variable $\Nmeas$ on the state produced by our ansatz with $\theta=\pi/4$ and extracting the minimum eigenvalue of the reconstructed density matrix (Fig.~\ref{fig:SOM_negativity}). A negative minimum eigenvalue manifests unphysicality over the $\Nmeas$ range covered. Our \emph{quantumsim} simulation produces a similar trend, asymptoting to a physical state by $\Nmeas=5\times10^4$. These observations led us to choose $\Nmeas=10^5$ for the final state tomography post VQE convergence in Fig.~\ref{fig:SV_impact_fig}, and to further investigate (in Fig.~\ref{fig:physicality}) how unphysicality can violate the variational principle, producing reductions in energy from imprecise state reconstruction rather than algorithmic precision.

\section{Constraining the positivity of reduced density matrices}~\label{app:positivity}
Testing whether a $N$-qubit density matrix $\rho$ is positive is in general QMA-hard~\cite{Liu07}.
However, if we trace out all but a polynomial number of degrees of freedom of $\rho$, testing positivity of the resulting reduced density matrix $\rho^{(\mathrm{red})}$ is tractable on classical hardware, and obtaining the closest nearby positive matrix is similarly so.
This gives a set of necessary but insufficient physicality conditions for $\rho$, but enforcing $k$-local constraints (on a density matrix from a VQE) tends to be sufficient to variationally bound the resulting energies~\cite{Rubin18}.
Following the reduction, we write $\rho^{(\mathrm{red})}$ as a vector over the Pauli basis,
\begin{equation}
\rho^{(\mathrm{red})}=\sum_{\hat{P}}\rho^{(\mathrm{red})}_{\hat{P}}\hat{P}.
\end{equation}
Then, we attempt to find the density matrix $\tilde{\rho}^{(\mathrm{red})}$ closest to $\rho^{(\mathrm{red})}$ in the $L^2$-norm
\begin{equation}
\sum_{\hat{P}}\left(\rho^{(\mathrm{red})}_{\hat{P}}-\tilde{\rho}^{(\mathrm{red})}_{\hat{P}}\right)^2,
\end{equation}
subject to the conditions $\tilde{\rho}^{(\mathrm{red})}\succ 0$, and $\tilde{\rho}^{(\mathrm{red})}_{I}=1$.
This gives a quadratic minimization problem with cone inequality and linear equality constraints, which we solve using interior point methods.

\section{Theoretical modeling of the experiment}
We use our full-density-matrix simulator \emph{quantumsim} to model the experiment.
The error model takes as input parameters the measured values of $\Tone$, $\Ttwos$ and residual excitation for both qubits, and $\Ttwosred$ for $\Qzero$.
We also include the effect of fluctuations on the device parameters by Monte Carlo sampling.

\subsection{Numerical simulations}
  The simulations are performed by extracting the full-density-matrix $\rhosim$ at the end of the circuit.
  We use the converged value of $\tilde{\theta}$ at each $R$ to generate the quantum state and extract the Pauli coefficients $\rhosimp(\tilde{\theta})=\trace[\hat{P}\rhosim]$. We add sampling noise to each coefficient, drawn from a zero-mean Gaussian distribution with variance $(1+\rhosimp)(1-\rhosimp)/\Nmeas$, where $\Nmeas=4\times10^5$. Note that this is greater than the number of measurements per tomographic prerotation in the experiment, as data from multiple tomographic prerotations is used to estimate each Pauli coefficient.
  To account for fluctuations on the device parameters $\Tone$, $\Ttwos$, and residual excitations, we average over $10^4$ simulations for every $R$.
  For each simulation, we draw parameters from independent normal distributions using values in Table~\ref{table:specs}.
  As the dephasing noise $\Ttwos$ depends on $\Tone$,
  \begin{equation}
  \frac{1}{\Ttwos} = \frac{1}{\Tphi} + \frac{1}{2 \Tone},
  \end{equation}
  it is more appropriate to sample the pure dephasing rate $\Tphi$ independently.
  We calculate the $\Tphi$ mean and variance ($\Tphim$, $\mathrm{Var}[\Tphim]$) from $\Ttwos$,
  \begin{equation}
  \begin{split}
  \Tphim = \frac{1}{\frac{1}{\Ttwos} - \frac{1}{2 \Tone}}, \\
  \mathrm{Var}[\Tphim] = \Tphim^{2} (\Ttwos)^{-2} \Bigg[\mathrm{Var}[\Ttwos] - \frac{\mathrm{Var}[\Tone]}{2 \Tone^{2} (\Ttwos)^{-2}}\Bigg].
  \end{split}
  \end{equation}
  From the $10^4$ simulations we obtain $95\%$-confidence error bars for $\Eerr$ and $\Fidel$ as twice the population standard deviation.

\subsection{Exchange gate}
Due to quasi-static flux noise, the angle of the unitary exchange gate (Eq.~\ref{eq:iswap_unitary}) differs between subsequent applications.
Assuming that these fluctuations are fast on the scale of the 2 hour optimization, this may be simulated by integrating over the range of applied gates, resulting in an incoherent noise model.
As the gate is not repeatedly applied during a single-shot experiment, this incoherent approximation does not lead to an error in the final result.
To perform the integration, we convert our unitary $\iSWAPtheta$ into a Pauli Transfer Matrix representation (PTM)~\cite{Chow12,Greenbaum}:
  \begin{equation}
  [R_\theta]_{i,j} = \frac{1}{2} \trace[\hat{P}_i \iSWAPtheta \hat{P}_j\iSWAPtheta^{\dag}],\hspace{0.5cm} \hat{P}_i,\hat{P}_j\in\pauligroup^N,
  \end{equation}
which may then be integrated over a probability distribution in the deviation $\delta$ from the target angle $\theta$:
  \begin{equation}
  [\tilde{R}]_{i,j} = \int{d \delta \ p(\delta) \ [R_{\theta+\delta}]_{i,j}}.
  \end{equation}
We choose for $p(\delta)$ a Gaussian distribution: $p(\delta) = e^{-\frac{\delta^2}{2 \sigma^2}}$.
In order to obtain the distribution width $\sigma^2$, we note that the same effect causes single-qubit dephasing of $\Qzero$ when fluxed to the exchange point when $\Qone$ is fluxed away.
We may thus estimate $\sigma$ as
  \begin{equation}
  \sigma^2 = 1 - e^{-\frac{\tinteract}{\Ttwosred}},
  \end{equation}
were $\tinteract$ is the exchange gate duration and $\Ttwosred$ the dephasing time of $\Qzero$ at the exchange point (with $\Qone$ fluxed away).
The final gate simulation also includes the effect of amplitude damping on both qubits, and the dephasing of $\Qone$ at the sweet spot as discrete error channels of duration $\frac{\tinteract}{2}$ on either side of the gate $\tilde{R}$. 

\begin{thebibliography}{49}%
\makeatletter
\providecommand \@ifxundefined [1]{%
 \@ifx{#1\undefined}
}%
\providecommand \@ifnum [1]{%
 \ifnum #1\expandafter \@firstoftwo
 \else \expandafter \@secondoftwo
 \fi
}%
\providecommand \@ifx [1]{%
 \ifx #1\expandafter \@firstoftwo
 \else \expandafter \@secondoftwo
 \fi
}%
\providecommand \natexlab [1]{#1}%
\providecommand \enquote  [1]{``#1''}%
\providecommand \bibnamefont  [1]{#1}%
\providecommand \bibfnamefont [1]{#1}%
\providecommand \citenamefont [1]{#1}%
\providecommand \href@noop [0]{\@secondoftwo}%
\providecommand \href [0]{\begingroup \@sanitize@url \@href}%
\providecommand \@href[1]{\@@startlink{#1}\@@href}%
\providecommand \@@href[1]{\endgroup#1\@@endlink}%
\providecommand \@sanitize@url [0]{\catcode `\\12\catcode `\$12\catcode
  `\&12\catcode `\#12\catcode `\^12\catcode `\_12\catcode `\%12\relax}%
\providecommand \@@startlink[1]{}%
\providecommand \@@endlink[0]{}%
\providecommand \url  [0]{\begingroup\@sanitize@url \@url }%
\providecommand \@url [1]{\endgroup\@href {#1}{\urlprefix }}%
\providecommand \urlprefix  [0]{URL }%
\providecommand \Eprint [0]{\href }%
\providecommand \doibase [0]{http://dx.doi.org/}%
\providecommand \selectlanguage [0]{\@gobble}%
\providecommand \bibinfo  [0]{\@secondoftwo}%
\providecommand \bibfield  [0]{\@secondoftwo}%
\providecommand \translation [1]{[#1]}%
\providecommand \BibitemOpen [0]{}%
\providecommand \bibitemStop [0]{}%
\providecommand \bibitemNoStop [0]{.\EOS\space}%
\providecommand \EOS [0]{\spacefactor3000\relax}%
\providecommand \BibitemShut  [1]{\csname bibitem#1\endcsname}%
\let\auto@bib@innerbib\@empty
\bibitem [{\citenamefont {Preskill}(2018)}]{Preskill2018NISQ}%
  \BibitemOpen
  \bibfield  {author} {\bibinfo {author} {\bibfnamefont {J.}~\bibnamefont
  {Preskill}},\ }\href {\doibase 10.22331/q-2018-08-06-79} {\bibfield
  {journal} {\bibinfo  {journal} {{Quantum}}\ }\textbf {\bibinfo {volume}
  {2}},\ \bibinfo {pages} {79} (\bibinfo {year} {2018})}\BibitemShut {NoStop}%
\bibitem [{\citenamefont {McArdle}\ \emph
  {et~al.}(2018{\natexlab{a}})\citenamefont {McArdle}, \citenamefont {Endo},
  \citenamefont {Aspuru-Guzik}, \citenamefont {Benjamin},\ and\ \citenamefont
  {Yuan}}]{Mca18Review}%
  \BibitemOpen
  \bibfield  {author} {\bibinfo {author} {\bibfnamefont {S.}~\bibnamefont
  {McArdle}}, \bibinfo {author} {\bibfnamefont {S.}~\bibnamefont {Endo}},
  \bibinfo {author} {\bibfnamefont {A.}~\bibnamefont {Aspuru-Guzik}}, \bibinfo
  {author} {\bibfnamefont {S.}~\bibnamefont {Benjamin}}, \ and\ \bibinfo
  {author} {\bibfnamefont {X.}~\bibnamefont {Yuan}},\ }\href
  {https://arxiv.org/abs/1808.10402} {\bibfield  {journal} {\bibinfo  {journal}
  {ArXiv:1808.10402}\ } (\bibinfo {year} {2018}{\natexlab{a}})}\BibitemShut
  {NoStop}%
\bibitem [{\citenamefont {Cao}\ \emph {et~al.}(2018)\citenamefont {Cao},
  \citenamefont {Romero}, \citenamefont {Olson}, \citenamefont {Degroote},
  \citenamefont {Johnson}, \citenamefont {Kieferov\'{a}}, \citenamefont
  {Kivlichan}, \citenamefont {Menke}, \citenamefont {Peropadre}, \citenamefont
  {Sawaya}, \citenamefont {Sim}, \citenamefont {Vies},\ and\ \citenamefont
  {Aspuru-Guzik}}]{Cao18Review}%
  \BibitemOpen
  \bibfield  {author} {\bibinfo {author} {\bibfnamefont {Y.}~\bibnamefont
  {Cao}}, \bibinfo {author} {\bibfnamefont {J.}~\bibnamefont {Romero}},
  \bibinfo {author} {\bibfnamefont {J.}~\bibnamefont {Olson}}, \bibinfo
  {author} {\bibfnamefont {M.}~\bibnamefont {Degroote}}, \bibinfo {author}
  {\bibfnamefont {P.}~\bibnamefont {Johnson}}, \bibinfo {author} {\bibfnamefont
  {M.}~\bibnamefont {Kieferov\'{a}}}, \bibinfo {author} {\bibfnamefont
  {I.}~\bibnamefont {Kivlichan}}, \bibinfo {author} {\bibfnamefont
  {T.}~\bibnamefont {Menke}}, \bibinfo {author} {\bibfnamefont
  {B.}~\bibnamefont {Peropadre}}, \bibinfo {author} {\bibfnamefont
  {N.}~\bibnamefont {Sawaya}}, \bibinfo {author} {\bibfnamefont
  {S.}~\bibnamefont {Sim}}, \bibinfo {author} {\bibfnamefont {L.}~\bibnamefont
  {Vies}}, \ and\ \bibinfo {author} {\bibfnamefont {A.}~\bibnamefont
  {Aspuru-Guzik}},\ }\href {https://arxiv.org/abs/1812.09976} {\bibfield
  {journal} {\bibinfo  {journal} {ArXiv:1812.09976}\ } (\bibinfo {year}
  {2018})}\BibitemShut {NoStop}%
\bibitem [{\citenamefont {Peruzzo}\ \emph {et~al.}(2014)\citenamefont
  {Peruzzo}, \citenamefont {McClean}, \citenamefont {Shadbolt}, \citenamefont
  {Yung}, \citenamefont {Zhou}, \citenamefont {Love}, \citenamefont
  {Aspuru-Guzik},\ and\ \citenamefont {O’Brien}}]{Peruzzo14}%
  \BibitemOpen
  \bibfield  {author} {\bibinfo {author} {\bibfnamefont {A.}~\bibnamefont
  {Peruzzo}}, \bibinfo {author} {\bibfnamefont {J.~R.}\ \bibnamefont
  {McClean}}, \bibinfo {author} {\bibfnamefont {P.}~\bibnamefont {Shadbolt}},
  \bibinfo {author} {\bibfnamefont {M.}~\bibnamefont {Yung}}, \bibinfo {author}
  {\bibfnamefont {X.}~\bibnamefont {Zhou}}, \bibinfo {author} {\bibfnamefont
  {P.~J.}\ \bibnamefont {Love}}, \bibinfo {author} {\bibfnamefont
  {A.}~\bibnamefont {Aspuru-Guzik}}, \ and\ \bibinfo {author} {\bibfnamefont
  {J.~L.}\ \bibnamefont {O’Brien}},\ }\href {\doibase 10.1038/ncomms5213}
  {\bibfield  {journal} {\bibinfo  {journal} {Nat.\ Commun.}\ }\textbf
  {\bibinfo {volume} {5}},\ \bibinfo {pages} {4213} (\bibinfo {year}
  {2014})}\BibitemShut {NoStop}%
\bibitem [{\citenamefont {McClean}\ \emph {et~al.}(2016)\citenamefont
  {McClean}, \citenamefont {Romero}, \citenamefont {Babbush},\ and\
  \citenamefont {Aspuru-Guzik}}]{Mcc16}%
  \BibitemOpen
  \bibfield  {author} {\bibinfo {author} {\bibfnamefont {J.~R.}\ \bibnamefont
  {McClean}}, \bibinfo {author} {\bibfnamefont {J.}~\bibnamefont {Romero}},
  \bibinfo {author} {\bibfnamefont {R.}~\bibnamefont {Babbush}}, \ and\
  \bibinfo {author} {\bibfnamefont {A.}~\bibnamefont {Aspuru-Guzik}},\ }\href
  {\doibase 10.1088/1367-2630/18/2/023023} {\bibfield  {journal} {\bibinfo
  {journal} {New Journal of Physics}\ }\textbf {\bibinfo {volume} {18}},\
  \bibinfo {pages} {023023} (\bibinfo {year} {2016})}\BibitemShut {NoStop}%
\bibitem [{\citenamefont {Babbush}\ \emph {et~al.}(2018)\citenamefont
  {Babbush}, \citenamefont {Wiebe}, \citenamefont {McClean}, \citenamefont
  {McClain}, \citenamefont {Neven},\ and\ \citenamefont {Chan}}]{Babbush18}%
  \BibitemOpen
  \bibfield  {author} {\bibinfo {author} {\bibfnamefont {R.}~\bibnamefont
  {Babbush}}, \bibinfo {author} {\bibfnamefont {N.}~\bibnamefont {Wiebe}},
  \bibinfo {author} {\bibfnamefont {J.}~\bibnamefont {McClean}}, \bibinfo
  {author} {\bibfnamefont {J.}~\bibnamefont {McClain}}, \bibinfo {author}
  {\bibfnamefont {H.}~\bibnamefont {Neven}}, \ and\ \bibinfo {author}
  {\bibfnamefont {G.~K.-L.}\ \bibnamefont {Chan}},\ }\href {\doibase
  10.1103/PhysRevX.8.011044} {\bibfield  {journal} {\bibinfo  {journal} {Phys.
  Rev. X}\ }\textbf {\bibinfo {volume} {8}},\ \bibinfo {pages} {011044}
  (\bibinfo {year} {2018})}\BibitemShut {NoStop}%
\bibitem [{\citenamefont {Poulin}\ \emph {et~al.}(2018)\citenamefont {Poulin},
  \citenamefont {Kitaev}, \citenamefont {Steiger}, \citenamefont {Hastings},\
  and\ \citenamefont {Troyer}}]{Poulin18}%
  \BibitemOpen
  \bibfield  {author} {\bibinfo {author} {\bibfnamefont {D.}~\bibnamefont
  {Poulin}}, \bibinfo {author} {\bibfnamefont {A.}~\bibnamefont {Kitaev}},
  \bibinfo {author} {\bibfnamefont {D.~S.}\ \bibnamefont {Steiger}}, \bibinfo
  {author} {\bibfnamefont {M.~B.}\ \bibnamefont {Hastings}}, \ and\ \bibinfo
  {author} {\bibfnamefont {M.}~\bibnamefont {Troyer}},\ }\href {\doibase
  10.1103/PhysRevLett.121.010501} {\bibfield  {journal} {\bibinfo  {journal}
  {Phys. Rev. Lett.}\ }\textbf {\bibinfo {volume} {121}},\ \bibinfo {pages}
  {010501} (\bibinfo {year} {2018})}\BibitemShut {NoStop}%
\bibitem [{\citenamefont {Berry}\ \emph {et~al.}(2018)\citenamefont {Berry},
  \citenamefont {Kieferov\'{a}}, \citenamefont {Scherer}, \citenamefont
  {Sanders}, \citenamefont {Low}, \citenamefont {Wiebe}, \citenamefont
  {Gidney},\ and\ \citenamefont {Babbush}}]{Berry18}%
  \BibitemOpen
  \bibfield  {author} {\bibinfo {author} {\bibfnamefont {D.~W.}\ \bibnamefont
  {Berry}}, \bibinfo {author} {\bibfnamefont {M.}~\bibnamefont
  {Kieferov\'{a}}}, \bibinfo {author} {\bibfnamefont {A.}~\bibnamefont
  {Scherer}}, \bibinfo {author} {\bibfnamefont {Y.~R.}\ \bibnamefont
  {Sanders}}, \bibinfo {author} {\bibfnamefont {G.~H.}\ \bibnamefont {Low}},
  \bibinfo {author} {\bibfnamefont {N.}~\bibnamefont {Wiebe}}, \bibinfo
  {author} {\bibfnamefont {C.}~\bibnamefont {Gidney}}, \ and\ \bibinfo {author}
  {\bibfnamefont {R.}~\bibnamefont {Babbush}},\ }\href {\doibase
  10.1038/s41534-018-0071-5} {\bibfield  {journal} {\bibinfo  {journal} {npj
  Quant. Inf.}\ }\textbf {\bibinfo {volume} {4}},\ \bibinfo {pages} {22}
  (\bibinfo {year} {2018})}\BibitemShut {NoStop}%
\bibitem [{\citenamefont {Kivlichan}\ \emph {et~al.}(2018)\citenamefont
  {Kivlichan}, \citenamefont {McClean}, \citenamefont {Wiebe}, \citenamefont
  {Gidney}, \citenamefont {Aspuru-Guzik}, \citenamefont {Chan},\ and\
  \citenamefont {Babbush}}]{Kivlichan18}%
  \BibitemOpen
  \bibfield  {author} {\bibinfo {author} {\bibfnamefont {I.~D.}\ \bibnamefont
  {Kivlichan}}, \bibinfo {author} {\bibfnamefont {J.}~\bibnamefont {McClean}},
  \bibinfo {author} {\bibfnamefont {N.}~\bibnamefont {Wiebe}}, \bibinfo
  {author} {\bibfnamefont {C.}~\bibnamefont {Gidney}}, \bibinfo {author}
  {\bibfnamefont {A.}~\bibnamefont {Aspuru-Guzik}}, \bibinfo {author}
  {\bibfnamefont {G.~K.-L.}\ \bibnamefont {Chan}}, \ and\ \bibinfo {author}
  {\bibfnamefont {R.}~\bibnamefont {Babbush}},\ }\href {\doibase
  10.1103/PhysRevLett.120.110501} {\bibfield  {journal} {\bibinfo  {journal}
  {Phys. Rev. Lett.}\ }\textbf {\bibinfo {volume} {120}},\ \bibinfo {pages}
  {110501} (\bibinfo {year} {2018})}\BibitemShut {NoStop}%
\bibitem [{\citenamefont {O'Malley}\ \emph {et~al.}(2016)\citenamefont
  {O'Malley}, \citenamefont {Babbush}, \citenamefont {Kivlichan}, \citenamefont
  {Romero}, \citenamefont {McClean}, \citenamefont {Barends}, \citenamefont
  {Kelly}, \citenamefont {Roushan}, \citenamefont {Tranter}, \citenamefont
  {Ding}, \citenamefont {Campbell}, \citenamefont {Chen}, \citenamefont {Chen},
  \citenamefont {Chiaro}, \citenamefont {Dunsworth}, \citenamefont {Fowler},
  \citenamefont {Jeffrey}, \citenamefont {Lucero}, \citenamefont {Megrant},
  \citenamefont {Mutus}, \citenamefont {Neeley}, \citenamefont {Neill},
  \citenamefont {Quintana}, \citenamefont {Sank}, \citenamefont {Vainsencher},
  \citenamefont {Wenner}, \citenamefont {White}, \citenamefont {Coveney},
  \citenamefont {Love}, \citenamefont {Neven}, \citenamefont {Aspuru-Guzik},\
  and\ \citenamefont {Martinis}}]{OMalley16}%
  \BibitemOpen
  \bibfield  {author} {\bibinfo {author} {\bibfnamefont {P.~J.~J.}\
  \bibnamefont {O'Malley}}, \bibinfo {author} {\bibfnamefont {R.}~\bibnamefont
  {Babbush}}, \bibinfo {author} {\bibfnamefont {I.~D.}\ \bibnamefont
  {Kivlichan}}, \bibinfo {author} {\bibfnamefont {J.}~\bibnamefont {Romero}},
  \bibinfo {author} {\bibfnamefont {J.~R.}\ \bibnamefont {McClean}}, \bibinfo
  {author} {\bibfnamefont {R.}~\bibnamefont {Barends}}, \bibinfo {author}
  {\bibfnamefont {J.}~\bibnamefont {Kelly}}, \bibinfo {author} {\bibfnamefont
  {P.}~\bibnamefont {Roushan}}, \bibinfo {author} {\bibfnamefont
  {A.}~\bibnamefont {Tranter}}, \bibinfo {author} {\bibfnamefont
  {N.}~\bibnamefont {Ding}}, \bibinfo {author} {\bibfnamefont {B.}~\bibnamefont
  {Campbell}}, \bibinfo {author} {\bibfnamefont {Y.}~\bibnamefont {Chen}},
  \bibinfo {author} {\bibfnamefont {Z.}~\bibnamefont {Chen}}, \bibinfo {author}
  {\bibfnamefont {B.}~\bibnamefont {Chiaro}}, \bibinfo {author} {\bibfnamefont
  {A.}~\bibnamefont {Dunsworth}}, \bibinfo {author} {\bibfnamefont {A.~G.}\
  \bibnamefont {Fowler}}, \bibinfo {author} {\bibfnamefont {E.}~\bibnamefont
  {Jeffrey}}, \bibinfo {author} {\bibfnamefont {E.}~\bibnamefont {Lucero}},
  \bibinfo {author} {\bibfnamefont {A.}~\bibnamefont {Megrant}}, \bibinfo
  {author} {\bibfnamefont {J.~Y.}\ \bibnamefont {Mutus}}, \bibinfo {author}
  {\bibfnamefont {M.}~\bibnamefont {Neeley}}, \bibinfo {author} {\bibfnamefont
  {C.}~\bibnamefont {Neill}}, \bibinfo {author} {\bibfnamefont
  {C.}~\bibnamefont {Quintana}}, \bibinfo {author} {\bibfnamefont
  {D.}~\bibnamefont {Sank}}, \bibinfo {author} {\bibfnamefont {A.}~\bibnamefont
  {Vainsencher}}, \bibinfo {author} {\bibfnamefont {J.}~\bibnamefont {Wenner}},
  \bibinfo {author} {\bibfnamefont {T.~C.}\ \bibnamefont {White}}, \bibinfo
  {author} {\bibfnamefont {P.~V.}\ \bibnamefont {Coveney}}, \bibinfo {author}
  {\bibfnamefont {P.~J.}\ \bibnamefont {Love}}, \bibinfo {author}
  {\bibfnamefont {H.}~\bibnamefont {Neven}}, \bibinfo {author} {\bibfnamefont
  {A.}~\bibnamefont {Aspuru-Guzik}}, \ and\ \bibinfo {author} {\bibfnamefont
  {J.~M.}\ \bibnamefont {Martinis}},\ }\href
  {https://link.aps.org/doi/10.1103/PhysRevX.6.031007} {\bibfield  {journal}
  {\bibinfo  {journal} {Phys. Rev. X}\ }\textbf {\bibinfo {volume} {6}},\
  \bibinfo {pages} {031007} (\bibinfo {year} {2016})}\BibitemShut {NoStop}%
\bibitem [{\citenamefont {Colless}\ \emph {et~al.}(2018)\citenamefont
  {Colless}, \citenamefont {Ramasesh}, \citenamefont {Dahlen}, \citenamefont
  {Blok}, \citenamefont {Kimchi-Schwartz}, \citenamefont {McClean},
  \citenamefont {Carter}, \citenamefont {de~Jong},\ and\ \citenamefont
  {Siddiqi}}]{Colless18}%
  \BibitemOpen
  \bibfield  {author} {\bibinfo {author} {\bibfnamefont {J.~I.}\ \bibnamefont
  {Colless}}, \bibinfo {author} {\bibfnamefont {V.~V.}\ \bibnamefont
  {Ramasesh}}, \bibinfo {author} {\bibfnamefont {D.}~\bibnamefont {Dahlen}},
  \bibinfo {author} {\bibfnamefont {M.~S.}\ \bibnamefont {Blok}}, \bibinfo
  {author} {\bibfnamefont {M.~E.}\ \bibnamefont {Kimchi-Schwartz}}, \bibinfo
  {author} {\bibfnamefont {J.~R.}\ \bibnamefont {McClean}}, \bibinfo {author}
  {\bibfnamefont {J.}~\bibnamefont {Carter}}, \bibinfo {author} {\bibfnamefont
  {W.~A.}\ \bibnamefont {de~Jong}}, \ and\ \bibinfo {author} {\bibfnamefont
  {I.}~\bibnamefont {Siddiqi}},\ }\href {\doibase 10.1103/PhysRevX.8.011021}
  {\bibfield  {journal} {\bibinfo  {journal} {Phys. Rev. X}\ }\textbf {\bibinfo
  {volume} {8}},\ \bibinfo {pages} {011021} (\bibinfo {year}
  {2018})}\BibitemShut {NoStop}%
\bibitem [{\citenamefont {Ganzhorn}\ \emph {et~al.}(2018)\citenamefont
  {Ganzhorn}, \citenamefont {Egger}, \citenamefont {Barkoutsos}, \citenamefont
  {Ollitrault}, \citenamefont {Salis}, \citenamefont {Moll}, \citenamefont
  {Fuhrer}, \citenamefont {Mueller}, \citenamefont {Woerner}, \citenamefont
  {Tavernelli},\ and\ \citenamefont {Filipp}}]{Ganzhorn18}%
  \BibitemOpen
  \bibfield  {author} {\bibinfo {author} {\bibfnamefont {M.}~\bibnamefont
  {Ganzhorn}}, \bibinfo {author} {\bibfnamefont {D.}~\bibnamefont {Egger}},
  \bibinfo {author} {\bibfnamefont {P.}~\bibnamefont {Barkoutsos}}, \bibinfo
  {author} {\bibfnamefont {P.}~\bibnamefont {Ollitrault}}, \bibinfo {author}
  {\bibfnamefont {G.}~\bibnamefont {Salis}}, \bibinfo {author} {\bibfnamefont
  {N.}~\bibnamefont {Moll}}, \bibinfo {author} {\bibfnamefont {A.}~\bibnamefont
  {Fuhrer}}, \bibinfo {author} {\bibfnamefont {P.}~\bibnamefont {Mueller}},
  \bibinfo {author} {\bibfnamefont {S.}~\bibnamefont {Woerner}}, \bibinfo
  {author} {\bibfnamefont {I.}~\bibnamefont {Tavernelli}}, \ and\ \bibinfo
  {author} {\bibfnamefont {S.}~\bibnamefont {Filipp}},\ }\href
  {https://arxiv.org/pdf/1809.05057.pdf} {\bibfield  {journal} {\bibinfo
  {journal} {arXiv: 1809.05057 [quant-ph]}\ } (\bibinfo {year}
  {2018})}\BibitemShut {NoStop}%
\bibitem [{\citenamefont {Hempel}\ \emph {et~al.}(2018)\citenamefont {Hempel},
  \citenamefont {Maier}, \citenamefont {Romero}, \citenamefont {McClean},
  \citenamefont {Monz}, \citenamefont {Shen}, \citenamefont {Jurcevic},
  \citenamefont {Lanyon}, \citenamefont {Love}, \citenamefont {Babbush},
  \citenamefont {Aspuru-Guzik}, \citenamefont {Blatt},\ and\ \citenamefont
  {Roos}}]{Hempel18}%
  \BibitemOpen
  \bibfield  {author} {\bibinfo {author} {\bibfnamefont {C.}~\bibnamefont
  {Hempel}}, \bibinfo {author} {\bibfnamefont {C.}~\bibnamefont {Maier}},
  \bibinfo {author} {\bibfnamefont {J.}~\bibnamefont {Romero}}, \bibinfo
  {author} {\bibfnamefont {J.}~\bibnamefont {McClean}}, \bibinfo {author}
  {\bibfnamefont {T.}~\bibnamefont {Monz}}, \bibinfo {author} {\bibfnamefont
  {H.}~\bibnamefont {Shen}}, \bibinfo {author} {\bibfnamefont {P.}~\bibnamefont
  {Jurcevic}}, \bibinfo {author} {\bibfnamefont {B.~P.}\ \bibnamefont
  {Lanyon}}, \bibinfo {author} {\bibfnamefont {P.}~\bibnamefont {Love}},
  \bibinfo {author} {\bibfnamefont {R.}~\bibnamefont {Babbush}}, \bibinfo
  {author} {\bibfnamefont {A.}~\bibnamefont {Aspuru-Guzik}}, \bibinfo {author}
  {\bibfnamefont {R.}~\bibnamefont {Blatt}}, \ and\ \bibinfo {author}
  {\bibfnamefont {C.~F.}\ \bibnamefont {Roos}},\ }\href {\doibase
  10.1103/PhysRevX.8.031022} {\bibfield  {journal} {\bibinfo  {journal} {Phys.
  Rev. X}\ }\textbf {\bibinfo {volume} {8}},\ \bibinfo {pages} {031022}
  (\bibinfo {year} {2018})}\BibitemShut {NoStop}%
\bibitem [{\citenamefont {Kandala}\ \emph {et~al.}(2017)\citenamefont
  {Kandala}, \citenamefont {Mezzacapo}, \citenamefont {Temme}, \citenamefont
  {Takita}, \citenamefont {Brink}, \citenamefont {Chow},\ and\ \citenamefont
  {Gambetta}}]{Kandala17}%
  \BibitemOpen
  \bibfield  {author} {\bibinfo {author} {\bibfnamefont {A.}~\bibnamefont
  {Kandala}}, \bibinfo {author} {\bibfnamefont {A.}~\bibnamefont {Mezzacapo}},
  \bibinfo {author} {\bibfnamefont {K.}~\bibnamefont {Temme}}, \bibinfo
  {author} {\bibfnamefont {M.}~\bibnamefont {Takita}}, \bibinfo {author}
  {\bibfnamefont {M.}~\bibnamefont {Brink}}, \bibinfo {author} {\bibfnamefont
  {J.~M.}\ \bibnamefont {Chow}}, \ and\ \bibinfo {author} {\bibfnamefont
  {J.~M.}\ \bibnamefont {Gambetta}},\ }\href {\doibase 10.1038/nature23879}
  {\bibfield  {journal} {\bibinfo  {journal} {Nature}\ }\textbf {\bibinfo
  {volume} {549}},\ \bibinfo {pages} {242} (\bibinfo {year}
  {2017})}\BibitemShut {NoStop}%
\bibitem [{\citenamefont {{Kandala}}\ \emph {et~al.}(2018)\citenamefont
  {{Kandala}}, \citenamefont {{Temme}}, \citenamefont {{Corcoles}},
  \citenamefont {{Mezzacapo}}, \citenamefont {{Chow}},\ and\ \citenamefont
  {{Gambetta}}}]{Kandala18}%
  \BibitemOpen
  \bibfield  {author} {\bibinfo {author} {\bibfnamefont {A.}~\bibnamefont
  {{Kandala}}}, \bibinfo {author} {\bibfnamefont {K.}~\bibnamefont {{Temme}}},
  \bibinfo {author} {\bibfnamefont {A.~D.}\ \bibnamefont {{Corcoles}}},
  \bibinfo {author} {\bibfnamefont {A.}~\bibnamefont {{Mezzacapo}}}, \bibinfo
  {author} {\bibfnamefont {J.~M.}\ \bibnamefont {{Chow}}}, \ and\ \bibinfo
  {author} {\bibfnamefont {J.~M.}\ \bibnamefont {{Gambetta}}},\ }\href
  {https://arxiv.org/abs/1805.04492} {\bibfield  {journal} {\bibinfo  {journal}
  {ArXiv:1805.04492}\ } (\bibinfo {year} {2018})}\BibitemShut {NoStop}%
\bibitem [{\citenamefont {Shen}\ \emph {et~al.}(2017)\citenamefont {Shen},
  \citenamefont {Zhang}, \citenamefont {Zhang}, \citenamefont {Zhang},
  \citenamefont {Yung},\ and\ \citenamefont {Kim}}]{Shen17}%
  \BibitemOpen
  \bibfield  {author} {\bibinfo {author} {\bibfnamefont {Y.}~\bibnamefont
  {Shen}}, \bibinfo {author} {\bibfnamefont {X.}~\bibnamefont {Zhang}},
  \bibinfo {author} {\bibfnamefont {S.}~\bibnamefont {Zhang}}, \bibinfo
  {author} {\bibfnamefont {J.-N.}\ \bibnamefont {Zhang}}, \bibinfo {author}
  {\bibfnamefont {M.-H.}\ \bibnamefont {Yung}}, \ and\ \bibinfo {author}
  {\bibfnamefont {K.}~\bibnamefont {Kim}},\ }\href
  {https://journals.aps.org/pra/abstract/10.1103/PhysRevA.95.020501} {\bibfield
   {journal} {\bibinfo  {journal} {Phys. Rev. A}\ }\textbf {\bibinfo {volume}
  {95}} (\bibinfo {year} {2017})}\BibitemShut {NoStop}%
\bibitem [{\citenamefont {Santagati}\ \emph {et~al.}(2018)\citenamefont
  {Santagati}, \citenamefont {Wang}, \citenamefont {Gentile}, \citenamefont
  {Paesani}, \citenamefont {Wiebe}, \citenamefont {McClean}, \citenamefont
  {Morley-Short}, \citenamefont {Shadbolt}, \citenamefont {Bonneau},
  \citenamefont {Silverstone}, \citenamefont {Tew}, \citenamefont {Zhou},
  \citenamefont {O'Brien},\ and\ \citenamefont {Thompson}}]{Santagati18}%
  \BibitemOpen
  \bibfield  {author} {\bibinfo {author} {\bibfnamefont {R.}~\bibnamefont
  {Santagati}}, \bibinfo {author} {\bibfnamefont {J.}~\bibnamefont {Wang}},
  \bibinfo {author} {\bibfnamefont {A.}~\bibnamefont {Gentile}}, \bibinfo
  {author} {\bibfnamefont {S.}~\bibnamefont {Paesani}}, \bibinfo {author}
  {\bibfnamefont {N.}~\bibnamefont {Wiebe}}, \bibinfo {author} {\bibfnamefont
  {J.}~\bibnamefont {McClean}}, \bibinfo {author} {\bibfnamefont
  {S.}~\bibnamefont {Morley-Short}}, \bibinfo {author} {\bibfnamefont
  {P.}~\bibnamefont {Shadbolt}}, \bibinfo {author} {\bibfnamefont
  {D.}~\bibnamefont {Bonneau}}, \bibinfo {author} {\bibfnamefont
  {J.}~\bibnamefont {Silverstone}}, \bibinfo {author} {\bibfnamefont
  {D.}~\bibnamefont {Tew}}, \bibinfo {author} {\bibfnamefont {X.}~\bibnamefont
  {Zhou}}, \bibinfo {author} {\bibfnamefont {J.}~\bibnamefont {O'Brien}}, \
  and\ \bibinfo {author} {\bibfnamefont {M.}~\bibnamefont {Thompson}},\ }\href
  {http://advances.sciencemag.org/content/4/1/eaap9646} {\bibfield  {journal}
  {\bibinfo  {journal} {Sci.\ Adv.}\ }\textbf {\bibinfo {volume} {4}},\
  \bibinfo {pages} {eaap9646} (\bibinfo {year} {2018})}\BibitemShut {NoStop}%
\bibitem [{\citenamefont {Kokail}\ \emph {et~al.}(2018)\citenamefont {Kokail},
  \citenamefont {Maier}, \citenamefont {van Bijnen}, \citenamefont {Brydges},
  \citenamefont {Joshi}, \citenamefont {Jurcevic}, \citenamefont {Muschik},
  \citenamefont {Silvi}, \citenamefont {Blatt}, \citenamefont {Roos},\ and\
  \citenamefont {Zoller}}]{Kokail18}%
  \BibitemOpen
  \bibfield  {author} {\bibinfo {author} {\bibfnamefont {C.}~\bibnamefont
  {Kokail}}, \bibinfo {author} {\bibfnamefont {C.}~\bibnamefont {Maier}},
  \bibinfo {author} {\bibfnamefont {R.}~\bibnamefont {van Bijnen}}, \bibinfo
  {author} {\bibfnamefont {T.}~\bibnamefont {Brydges}}, \bibinfo {author}
  {\bibfnamefont {M.}~\bibnamefont {Joshi}}, \bibinfo {author} {\bibfnamefont
  {P.}~\bibnamefont {Jurcevic}}, \bibinfo {author} {\bibfnamefont
  {C.}~\bibnamefont {Muschik}}, \bibinfo {author} {\bibfnamefont
  {P.}~\bibnamefont {Silvi}}, \bibinfo {author} {\bibfnamefont
  {R.}~\bibnamefont {Blatt}}, \bibinfo {author} {\bibfnamefont
  {C.}~\bibnamefont {Roos}}, \ and\ \bibinfo {author} {\bibfnamefont
  {P.}~\bibnamefont {Zoller}},\ }\href {https://arxiv.org/abs/1810.03421}
  {\bibfield  {journal} {\bibinfo  {journal} {ArXiv:1810.03421}\ } (\bibinfo
  {year} {2018})}\BibitemShut {NoStop}%
\bibitem [{\citenamefont {Bonet-Monroig}\ \emph {et~al.}(2018)\citenamefont
  {Bonet-Monroig}, \citenamefont {Sagastizabal}, \citenamefont {Singh},\ and\
  \citenamefont {O'Brien}}]{Bonet18}%
  \BibitemOpen
  \bibfield  {author} {\bibinfo {author} {\bibfnamefont {X.}~\bibnamefont
  {Bonet-Monroig}}, \bibinfo {author} {\bibfnamefont {R.}~\bibnamefont
  {Sagastizabal}}, \bibinfo {author} {\bibfnamefont {M.}~\bibnamefont {Singh}},
  \ and\ \bibinfo {author} {\bibfnamefont {T.~E.}\ \bibnamefont {O'Brien}},\
  }\href {\doibase 10.1103/PhysRevA.98.062339} {\bibfield  {journal} {\bibinfo
  {journal} {Phys. Rev. A}\ }\textbf {\bibinfo {volume} {98}},\ \bibinfo
  {pages} {062339} (\bibinfo {year} {2018})}\BibitemShut {NoStop}%
\bibitem [{\citenamefont {McArdle}\ \emph
  {et~al.}(2018{\natexlab{b}})\citenamefont {McArdle}, \citenamefont {Yuan},\
  and\ \citenamefont {Benjamin}}]{Mca18}%
  \BibitemOpen
  \bibfield  {author} {\bibinfo {author} {\bibfnamefont {S.}~\bibnamefont
  {McArdle}}, \bibinfo {author} {\bibfnamefont {X.}~\bibnamefont {Yuan}}, \
  and\ \bibinfo {author} {\bibfnamefont {S.~C.}\ \bibnamefont {Benjamin}},\
  }\href {https://arxiv.org/abs/1807.02467} {\bibfield  {journal} {\bibinfo
  {journal} {ArXiv:1807.02467}\ } (\bibinfo {year}
  {2018}{\natexlab{b}})}\BibitemShut {NoStop}%
\bibitem [{\citenamefont {McClean}\ \emph {et~al.}(2017)\citenamefont
  {McClean}, \citenamefont {Kimchi-Schwartz}, \citenamefont {Carter},\ and\
  \citenamefont {de~Jong}}]{Mcc17}%
  \BibitemOpen
  \bibfield  {author} {\bibinfo {author} {\bibfnamefont {J.~R.}\ \bibnamefont
  {McClean}}, \bibinfo {author} {\bibfnamefont {M.~E.}\ \bibnamefont
  {Kimchi-Schwartz}}, \bibinfo {author} {\bibfnamefont {J.}~\bibnamefont
  {Carter}}, \ and\ \bibinfo {author} {\bibfnamefont {W.~A.}\ \bibnamefont
  {de~Jong}},\ }\href {\doibase 10.1103/PhysRevA.95.042308} {\bibfield
  {journal} {\bibinfo  {journal} {Phys. Rev. A}\ }\textbf {\bibinfo {volume}
  {95}},\ \bibinfo {pages} {042308} (\bibinfo {year} {2017})}\BibitemShut
  {NoStop}%
\bibitem [{\citenamefont {Li}\ \emph {et~al.}(2017)\citenamefont {Li},
  \citenamefont {Petit}, \citenamefont {Franke}, \citenamefont {Dehollain},
  \citenamefont {Helsen}, \citenamefont {Steudtner}, \citenamefont {Thomas},
  \citenamefont {Yoscovits}, \citenamefont {Singh}, \citenamefont {Wehner}
  \emph {et~al.}}]{Li17}%
  \BibitemOpen
  \bibfield  {author} {\bibinfo {author} {\bibfnamefont {R.}~\bibnamefont
  {Li}}, \bibinfo {author} {\bibfnamefont {L.}~\bibnamefont {Petit}}, \bibinfo
  {author} {\bibfnamefont {D.}~\bibnamefont {Franke}}, \bibinfo {author}
  {\bibfnamefont {J.}~\bibnamefont {Dehollain}}, \bibinfo {author}
  {\bibfnamefont {J.}~\bibnamefont {Helsen}}, \bibinfo {author} {\bibfnamefont
  {M.}~\bibnamefont {Steudtner}}, \bibinfo {author} {\bibfnamefont
  {N.}~\bibnamefont {Thomas}}, \bibinfo {author} {\bibfnamefont
  {Z.}~\bibnamefont {Yoscovits}}, \bibinfo {author} {\bibfnamefont
  {K.}~\bibnamefont {Singh}}, \bibinfo {author} {\bibfnamefont
  {S.}~\bibnamefont {Wehner}},  \emph {et~al.},\ }\href
  {https://arxiv.org/abs/1711.03807} {\bibfield  {journal} {\bibinfo  {journal}
  {arXiv:1711.03807}\ } (\bibinfo {year} {2017})}\BibitemShut {NoStop}%
\bibitem [{\citenamefont {Temme}\ \emph {et~al.}(2017)\citenamefont {Temme},
  \citenamefont {Bravyi},\ and\ \citenamefont {Gambetta}}]{Temme17}%
  \BibitemOpen
  \bibfield  {author} {\bibinfo {author} {\bibfnamefont {K.}~\bibnamefont
  {Temme}}, \bibinfo {author} {\bibfnamefont {S.}~\bibnamefont {Bravyi}}, \
  and\ \bibinfo {author} {\bibfnamefont {J.~M.}\ \bibnamefont {Gambetta}},\
  }\href {\doibase 10.1103/PhysRevLett.119.180509} {\bibfield  {journal}
  {\bibinfo  {journal} {Phys. Rev. Lett.}\ }\textbf {\bibinfo {volume} {119}},\
  \bibinfo {pages} {180509} (\bibinfo {year} {2017})}\BibitemShut {NoStop}%
\bibitem [{\citenamefont {Otten}\ and\ \citenamefont {Gray}(2018)}]{Otten18}%
  \BibitemOpen
  \bibfield  {author} {\bibinfo {author} {\bibfnamefont {M.}~\bibnamefont
  {Otten}}\ and\ \bibinfo {author} {\bibfnamefont {S.}~\bibnamefont {Gray}},\
  }\href {https://journals.aps.org/pra/abstract/10.1103/PhysRevA.99.012338}
  {\bibfield  {journal} {\bibinfo  {journal} {Phys. Rev. A}\ }\textbf {\bibinfo
  {volume} {99}} (\bibinfo {year} {2018})}\BibitemShut {NoStop}%
\bibitem [{\citenamefont {Endo}\ \emph {et~al.}(2018)\citenamefont {Endo},
  \citenamefont {Benjamin},\ and\ \citenamefont {Li}}]{Endo18}%
  \BibitemOpen
  \bibfield  {author} {\bibinfo {author} {\bibfnamefont {S.}~\bibnamefont
  {Endo}}, \bibinfo {author} {\bibfnamefont {S.}~\bibnamefont {Benjamin}}, \
  and\ \bibinfo {author} {\bibfnamefont {Y.}~\bibnamefont {Li}},\ }\href
  {https://journals.aps.org/prx/abstract/10.1103/PhysRevX.8.031027} {\bibfield
  {journal} {\bibinfo  {journal} {Phys. Rev. X}\ }\textbf {\bibinfo {volume}
  {8}} (\bibinfo {year} {2018})}\BibitemShut {NoStop}%
\bibitem [{\citenamefont {Huo}\ and\ \citenamefont {Li}(2018)}]{Huo18}%
  \BibitemOpen
  \bibfield  {author} {\bibinfo {author} {\bibfnamefont {M.}~\bibnamefont
  {Huo}}\ and\ \bibinfo {author} {\bibfnamefont {Y.}~\bibnamefont {Li}},\
  }\href {https://arxiv.org/abs/1811.02734} {\bibfield  {journal} {\bibinfo
  {journal} {ArXiv:1811.02734}\ } (\bibinfo {year} {2018})}\BibitemShut
  {NoStop}%
\bibitem [{\citenamefont {Bravyi}\ and\ \citenamefont
  {Kitaev}(2002)}]{Bravyi02BKmap}%
  \BibitemOpen
  \bibfield  {author} {\bibinfo {author} {\bibfnamefont {S.}~\bibnamefont
  {Bravyi}}\ and\ \bibinfo {author} {\bibfnamefont {A.}~\bibnamefont
  {Kitaev}},\ }\href {https://arxiv.org/abs/quant-ph/0003137} {\bibfield
  {journal} {\bibinfo  {journal} {ArXiv:quant-ph/0003137}\ } (\bibinfo {year}
  {2002})}\BibitemShut {NoStop}%
\bibitem [{\citenamefont {{McClean}}\ \emph {et~al.}()\citenamefont
  {{McClean}}, \citenamefont {{Kivlichan}}, \citenamefont {{Sung}},
  \citenamefont {{Steiger}}, \citenamefont {{Cao}}, \citenamefont {{Dai}},
  \citenamefont {{Schuyler Fried}}, \citenamefont {{Gidney}}, \citenamefont
  {{Gimby}}, \citenamefont {{Gokhale}}, \citenamefont {{H{\"a}ner}},
  \citenamefont {{Hardikar}}, \citenamefont {{Havl{\'{\i}}{\v c}ek}},
  \citenamefont {{Huang}}, \citenamefont {{Izaac}}, \citenamefont {{Jiang}},
  \citenamefont {{Liu}}, \citenamefont {{Neeley}}, \citenamefont {{O'Brien}},
  \citenamefont {{Ozfidan}}, \citenamefont {{Radin}}, \citenamefont {{Romero}},
  \citenamefont {{Rubin}}, \citenamefont {{Sawaya}}, \citenamefont {{Setia}},
  \citenamefont {{Sim}}, \citenamefont {{Steudtner}}, \citenamefont {{Sun}},
  \citenamefont {{Sun}}, \citenamefont {{Zhang}},\ and\ \citenamefont
  {{Babbush}}}]{Openfermion}%
  \BibitemOpen
  \bibfield  {author} {\bibinfo {author} {\bibfnamefont {J.~R.}\ \bibnamefont
  {{McClean}}}, \bibinfo {author} {\bibfnamefont {I.~D.}\ \bibnamefont
  {{Kivlichan}}}, \bibinfo {author} {\bibfnamefont {K.~J.}\ \bibnamefont
  {{Sung}}}, \bibinfo {author} {\bibfnamefont {D.~S.}\ \bibnamefont
  {{Steiger}}}, \bibinfo {author} {\bibfnamefont {Y.}~\bibnamefont {{Cao}}},
  \bibinfo {author} {\bibfnamefont {C.}~\bibnamefont {{Dai}}}, \bibinfo
  {author} {\bibfnamefont {E.}~\bibnamefont {{Schuyler Fried}}}, \bibinfo
  {author} {\bibfnamefont {C.}~\bibnamefont {{Gidney}}}, \bibinfo {author}
  {\bibfnamefont {B.}~\bibnamefont {{Gimby}}}, \bibinfo {author} {\bibfnamefont
  {P.}~\bibnamefont {{Gokhale}}}, \bibinfo {author} {\bibfnamefont
  {T.}~\bibnamefont {{H{\"a}ner}}}, \bibinfo {author} {\bibfnamefont
  {T.}~\bibnamefont {{Hardikar}}}, \bibinfo {author} {\bibfnamefont
  {V.}~\bibnamefont {{Havl{\'{\i}}{\v c}ek}}}, \bibinfo {author} {\bibfnamefont
  {C.}~\bibnamefont {{Huang}}}, \bibinfo {author} {\bibfnamefont
  {J.}~\bibnamefont {{Izaac}}}, \bibinfo {author} {\bibfnamefont
  {Z.}~\bibnamefont {{Jiang}}}, \bibinfo {author} {\bibfnamefont
  {X.}~\bibnamefont {{Liu}}}, \bibinfo {author} {\bibfnamefont
  {M.}~\bibnamefont {{Neeley}}}, \bibinfo {author} {\bibfnamefont
  {T.~E.}~\bibnamefont {{O'Brien}}}, \bibinfo {author} {\bibfnamefont
  {I.}~\bibnamefont {{Ozfidan}}}, \bibinfo {author} {\bibfnamefont {M.~D.}\
  \bibnamefont {{Radin}}}, \bibinfo {author} {\bibfnamefont {J.}~\bibnamefont
  {{Romero}}}, \bibinfo {author} {\bibfnamefont {N.}~\bibnamefont {{Rubin}}},
  \bibinfo {author} {\bibfnamefont {N.~P.~D.}\ \bibnamefont {{Sawaya}}},
  \bibinfo {author} {\bibfnamefont {K.}~\bibnamefont {{Setia}}}, \bibinfo
  {author} {\bibfnamefont {S.}~\bibnamefont {{Sim}}}, \bibinfo {author}
  {\bibfnamefont {M.}~\bibnamefont {{Steudtner}}}, \bibinfo {author}
  {\bibfnamefont {Q.}~\bibnamefont {{Sun}}}, \bibinfo {author} {\bibfnamefont
  {W.}~\bibnamefont {{Sun}}}, \bibinfo {author} {\bibfnamefont
  {F.}~\bibnamefont {{Zhang}}}, \ and\ \bibinfo {author} {\bibfnamefont
  {R.}~\bibnamefont {{Babbush}}},\ }\href {https://arxiv.org/abs/1710.07629}
  {\bibinfo  {journal} {ArXiv:1710.07629}\ }\BibitemShut {NoStop}%
\bibitem [{\citenamefont {Parrish}\ \emph {et~al.}(2017)\citenamefont
  {Parrish}, \citenamefont {Burns}, \citenamefont {Smith}, \citenamefont
  {Simmonett}, \citenamefont {DePrince}, \citenamefont {Hohenstein},
  \citenamefont {Bozkaya}, \citenamefont {Sokolov}, \citenamefont {Di~Remigio},
  \citenamefont {Richard}, \citenamefont {Gonthier}, \citenamefont {James},
  \citenamefont {McAlexander}, \citenamefont {Kumar}, \citenamefont {Saitow},
  \citenamefont {Wang}, \citenamefont {Pritchard}, \citenamefont {Verma},
  \citenamefont {Schaefer}, \citenamefont {Patkowski}, \citenamefont {King},
  \citenamefont {Valeev}, \citenamefont {Evangelista}, \citenamefont {Turney},
  \citenamefont {Crawford}, ,\ and\ \citenamefont {Sherrill}}]{Parrish17Psi4}%
  \BibitemOpen
\bibfield  {journal} {  }\bibfield  {author} {\bibinfo {author} {\bibfnamefont
  {R.~M.}\ \bibnamefont {Parrish}}, \bibinfo {author} {\bibfnamefont {L.~A.}\
  \bibnamefont {Burns}}, \bibinfo {author} {\bibfnamefont {D.~G.~A.}\
  \bibnamefont {Smith}}, \bibinfo {author} {\bibfnamefont {A.~C.}\ \bibnamefont
  {Simmonett}}, \bibinfo {author} {\bibfnamefont {A.~E.}\ \bibnamefont
  {DePrince}}, \bibinfo {author} {\bibfnamefont {E.~G.}\ \bibnamefont
  {Hohenstein}}, \bibinfo {author} {\bibfnamefont {U.}~\bibnamefont {Bozkaya}},
  \bibinfo {author} {\bibfnamefont {A.~Y.}\ \bibnamefont {Sokolov}}, \bibinfo
  {author} {\bibfnamefont {R.}~\bibnamefont {Di~Remigio}}, \bibinfo {author}
  {\bibfnamefont {R.~M.}\ \bibnamefont {Richard}}, \bibinfo {author}
  {\bibfnamefont {J.~F.}\ \bibnamefont {Gonthier}}, \bibinfo {author}
  {\bibfnamefont {A.~M.}\ \bibnamefont {James}}, \bibinfo {author}
  {\bibfnamefont {H.~R.}\ \bibnamefont {McAlexander}}, \bibinfo {author}
  {\bibfnamefont {A.}~\bibnamefont {Kumar}}, \bibinfo {author} {\bibfnamefont
  {M.}~\bibnamefont {Saitow}}, \bibinfo {author} {\bibfnamefont
  {X.}~\bibnamefont {Wang}}, \bibinfo {author} {\bibfnamefont {N.~P.}\
  \bibnamefont {Pritchard}}, \bibinfo {author} {\bibfnamefont {P.}~\bibnamefont
  {Verma}}, \bibinfo {author} {\bibfnamefont {H.~F.}\ \bibnamefont {Schaefer}},
  \bibinfo {author} {\bibfnamefont {K.}~\bibnamefont {Patkowski}}, \bibinfo
  {author} {\bibfnamefont {R.~A.}\ \bibnamefont {King}}, \bibinfo {author}
  {\bibfnamefont {E.~F.}\ \bibnamefont {Valeev}}, \bibinfo {author}
  {\bibfnamefont {F.~A.}\ \bibnamefont {Evangelista}}, \bibinfo {author}
  {\bibfnamefont {J.~M.}\ \bibnamefont {Turney}}, \bibinfo {author}
  {\bibfnamefont {T.~D.}\ \bibnamefont {Crawford}}, , \ and\ \bibinfo {author}
  {\bibfnamefont {C.~D.}\ \bibnamefont {Sherrill}},\ }\href {\doibase
  10.1021/acs.jctc.7b00174} {\bibfield  {journal} {\bibinfo  {journal} {Journal
  of Chemical Theory and Computation}\ }\textbf {\bibinfo {volume} {13}},\
  \bibinfo {pages} {3185} (\bibinfo {year} {2017})}\BibitemShut {NoStop}%
\bibitem [{Note1()}]{Note1}%
  \BibitemOpen
  \bibinfo {note} {As described in Refs.~\cite {Bonet18,Mca18}, one does not
  require $\protect \mathaccentV {hat}05E{S}$ to be a Pauli operator, however
  this makes the SV procedure significantly simpler.}\BibitemShut {Stop}%
\bibitem [{\citenamefont {McClean}\ \emph {et~al.}(2018)\citenamefont
  {McClean}, \citenamefont {Boixo}, \citenamefont {Smelyanskiy}, \citenamefont
  {Babbush},\ and\ \citenamefont {Neven}}]{McCleanplat18}%
  \BibitemOpen
  \bibfield  {author} {\bibinfo {author} {\bibfnamefont {J.}~\bibnamefont
  {McClean}}, \bibinfo {author} {\bibfnamefont {S.}~\bibnamefont {Boixo}},
  \bibinfo {author} {\bibfnamefont {V.}~\bibnamefont {Smelyanskiy}}, \bibinfo
  {author} {\bibfnamefont {R.}~\bibnamefont {Babbush}}, \ and\ \bibinfo
  {author} {\bibfnamefont {H.}~\bibnamefont {Neven}},\ }\href {\doibase
  10.1038/s41467-018-07090-4} {\bibfield  {journal} {\bibinfo  {journal} {Nat.\
  Commun.}\ }\textbf {\bibinfo {volume} {9}},\ \bibinfo {pages} {4812}
  (\bibinfo {year} {2018})}\BibitemShut {NoStop}%
\bibitem [{\citenamefont {Di{C}arlo}\ \emph {et~al.}(2009)\citenamefont
  {Di{C}arlo}, \citenamefont {Chow}, \citenamefont {Gambetta}, \citenamefont
  {Bishop}, \citenamefont {Johnson}, \citenamefont {Schuster}, \citenamefont
  {Majer}, \citenamefont {Blais}, \citenamefont {Frunzio}, \citenamefont
  {Girvin},\ and\ \citenamefont {Schoelkopf}}]{DiCarlo09}%
  \BibitemOpen
  \bibfield  {author} {\bibinfo {author} {\bibfnamefont {L.}~\bibnamefont
  {Di{C}arlo}}, \bibinfo {author} {\bibfnamefont {J.~M.}\ \bibnamefont {Chow}},
  \bibinfo {author} {\bibfnamefont {J.~M.}\ \bibnamefont {Gambetta}}, \bibinfo
  {author} {\bibfnamefont {L.~S.}\ \bibnamefont {Bishop}}, \bibinfo {author}
  {\bibfnamefont {B.~R.}\ \bibnamefont {Johnson}}, \bibinfo {author}
  {\bibfnamefont {D.~I.}\ \bibnamefont {Schuster}}, \bibinfo {author}
  {\bibfnamefont {J.}~\bibnamefont {Majer}}, \bibinfo {author} {\bibfnamefont
  {A.}~\bibnamefont {Blais}}, \bibinfo {author} {\bibfnamefont
  {L.}~\bibnamefont {Frunzio}}, \bibinfo {author} {\bibfnamefont {S.~M.}\
  \bibnamefont {Girvin}}, \ and\ \bibinfo {author} {\bibfnamefont {R.~J.}\
  \bibnamefont {Schoelkopf}},\ }\href
  {http://www.nature.com/nature/journal/v460/n7252/abs/nature08121.html}
  {\bibfield  {journal} {\bibinfo  {journal} {Nature}\ }\textbf {\bibinfo
  {volume} {460}},\ \bibinfo {pages} {240} (\bibinfo {year}
  {2009})}\BibitemShut {NoStop}%
\bibitem [{\citenamefont {Majer}\ \emph {et~al.}(2007)\citenamefont {Majer},
  \citenamefont {Chow}, \citenamefont {Gambetta}, \citenamefont {Johnson},
  \citenamefont {Schreier}, \citenamefont {Frunzio}, \citenamefont {Schuster},
  \citenamefont {Houck}, \citenamefont {Wallraff}, \citenamefont {Blais},
  \citenamefont {Devoret}, \citenamefont {Girvin},\ and\ \citenamefont
  {Schoelkopf}}]{Majer07}%
  \BibitemOpen
  \bibfield  {author} {\bibinfo {author} {\bibfnamefont {J.}~\bibnamefont
  {Majer}}, \bibinfo {author} {\bibfnamefont {J.~M.}\ \bibnamefont {Chow}},
  \bibinfo {author} {\bibfnamefont {J.~M.}\ \bibnamefont {Gambetta}}, \bibinfo
  {author} {\bibfnamefont {B.~R.}\ \bibnamefont {Johnson}}, \bibinfo {author}
  {\bibfnamefont {J.~A.}\ \bibnamefont {Schreier}}, \bibinfo {author}
  {\bibfnamefont {L.}~\bibnamefont {Frunzio}}, \bibinfo {author} {\bibfnamefont
  {D.~I.}\ \bibnamefont {Schuster}}, \bibinfo {author} {\bibfnamefont {A.~A.}\
  \bibnamefont {Houck}}, \bibinfo {author} {\bibfnamefont {A.}~\bibnamefont
  {Wallraff}}, \bibinfo {author} {\bibfnamefont {A.}~\bibnamefont {Blais}},
  \bibinfo {author} {\bibfnamefont {M.~H.}\ \bibnamefont {Devoret}}, \bibinfo
  {author} {\bibfnamefont {S.~M.}\ \bibnamefont {Girvin}}, \ and\ \bibinfo
  {author} {\bibfnamefont {R.~J.}\ \bibnamefont {Schoelkopf}},\ }\href{https://www.nature.com/articles/nature06184}
  {\bibfield  {journal} {\bibinfo  {journal} {Nature}\ }\textbf {\bibinfo
  {volume} {449}},\ \bibinfo {pages} {443} (\bibinfo {year}
  {2007})}\BibitemShut {NoStop}%
\bibitem [{\citenamefont {Brod}\ and\ \citenamefont {Childs}(2014)}]{Brod14}%
  \BibitemOpen
  \bibfield  {author} {\bibinfo {author} {\bibfnamefont {D.}~\bibnamefont
  {Brod}}\ and\ \bibinfo {author} {\bibfnamefont {A.}~\bibnamefont {Childs}},\
  }\href {http://www.rintonpress.com/journals/qiconline.html#v14n1112}
  {\bibfield  {journal} {\bibinfo  {journal} {Quantum Information \&
  Computation}\ }\textbf {\bibinfo {volume} {14}},\ \bibinfo {pages} {901}
  (\bibinfo {year} {2014})}\BibitemShut {NoStop}%
\bibitem [{SOM()}]{SOM_VQE}%
  \BibitemOpen
  \href@noop {} {}\bibinfo {howpublished} {See supplemental material at [insert
  URL] for additional data.}\BibitemShut {Stop}%
\bibitem [{\citenamefont {Hansen}(2009)}]{Hansen09}%
  \BibitemOpen
  \bibfield  {author} {\bibinfo {author} {\bibfnamefont {N.}~\bibnamefont
  {Hansen}},\ }in\ \href {\doibase 10.1145/1570256.1570335} {\emph {\bibinfo
  {booktitle} {Proceedings of the 11th Annual Conference Companion on Genetic
  and Evolutionary Computation Conference: Late Breaking Papers}}},\ \bibinfo
  {series and number} {GECCO '09}\ (\bibinfo  {publisher} {ACM},\ \bibinfo
  {address} {New York, NY, USA},\ \bibinfo {year} {2009})\ pp.\ \bibinfo
  {pages} {2403--2408}\BibitemShut {NoStop}%
\bibitem [{\citenamefont {O'Brien}\ \emph {et~al.}(2017)\citenamefont
  {O'Brien}, \citenamefont {Tarasinski},\ and\ \citenamefont
  {DiCarlo}}]{Obrien17}%
  \BibitemOpen
  \bibfield  {author} {\bibinfo {author} {\bibfnamefont {T.~E.}~\bibnamefont
  {O'Brien}}, \bibinfo {author} {\bibfnamefont {B.}~\bibnamefont {Tarasinski}},
  \ and\ \bibinfo {author} {\bibfnamefont {L.}~\bibnamefont {DiCarlo}},\ }\href
  {https://iopscience.iop.org/article/10.1088/1367-2630/aafb8e/pdf} {\bibfield
  {journal} {\bibinfo  {journal} {npj Quantum Information}\ }\textbf {\bibinfo
  {volume} {3}} (\bibinfo {year} {2017})}\BibitemShut {NoStop}%
\bibitem [{\citenamefont {Rubin}\ \emph {et~al.}(2018)\citenamefont {Rubin},
  \citenamefont {Babbush},\ and\ \citenamefont {McClean}}]{Rubin18}%
  \BibitemOpen
  \bibfield  {author} {\bibinfo {author} {\bibfnamefont {N.~C.}\ \bibnamefont
  {Rubin}}, \bibinfo {author} {\bibfnamefont {R.}~\bibnamefont {Babbush}}, \
  and\ \bibinfo {author} {\bibfnamefont {J.}~\bibnamefont {McClean}},\ }\href
  {\doibase 10.1088/1367-2630/aab919} {\bibfield  {journal} {\bibinfo
  {journal} {New J.\ Phys.}\ }\textbf {\bibinfo {volume} {20}},\ \bibinfo
  {pages} {053020} (\bibinfo {year} {2018})}\BibitemShut {NoStop}%
\bibitem [{\citenamefont {Blume-Kohout}(2010)}]{Blume06}%
  \BibitemOpen
  \bibfield  {author} {\bibinfo {author} {\bibfnamefont {R.}~\bibnamefont
  {Blume-Kohout}},\ }\href {\doibase 10.1088/1367-2630/12/4/043034} {\bibfield
  {journal} {\bibinfo  {journal} {New J.\ Phys.}\ }\textbf {\bibinfo {volume}
  {12}},\ \bibinfo {pages} {043034} (\bibinfo {year} {2010})}\BibitemShut
  {NoStop}%
\bibitem [{\citenamefont {Liu}\ \emph {et~al.}(2007)\citenamefont {Liu},
  \citenamefont {Christandl},\ and\ \citenamefont {Verstraete}}]{Liu07}%
  \BibitemOpen
  \bibfield  {author} {\bibinfo {author} {\bibfnamefont {Y.-K.}\ \bibnamefont
  {Liu}}, \bibinfo {author} {\bibfnamefont {M.}~\bibnamefont {Christandl}}, \
  and\ \bibinfo {author} {\bibfnamefont {F.}~\bibnamefont {Verstraete}},\
  }\href {\doibase 10.1103/PhysRevLett.98.110503} {\bibfield  {journal}
  {\bibinfo  {journal} {Phys. Rev. Lett.}\ }\textbf {\bibinfo {volume} {98}},\
  \bibinfo {pages} {110503} (\bibinfo {year} {2007})}\BibitemShut {NoStop}%
\bibitem [{Note2()}]{Note2}%
  \BibitemOpen
  \bibinfo {note} {Note that, for this system, enforcing positivity of the
  $1$-reduced density matrix corresponds to ensuring that all expectation
  values are bounded between $-1$ and $1$, and so this does not provide any
  additional data.}\BibitemShut {Stop}%
\bibitem [{\citenamefont {Macklin}\ \emph {et~al.}(2015)\citenamefont
  {Macklin}, \citenamefont {O{\textquoteright}Brien}, \citenamefont {Hover},
  \citenamefont {Schwartz}, \citenamefont {Bolkhovsky}, \citenamefont {Zhang},
  \citenamefont {Oliver},\ and\ \citenamefont {Siddiqi}}]{Macklin307}%
  \BibitemOpen
  \bibfield  {author} {\bibinfo {author} {\bibfnamefont {C.}~\bibnamefont
  {Macklin}}, \bibinfo {author} {\bibfnamefont {K.}~\bibnamefont
  {O{\textquoteright}Brien}}, \bibinfo {author} {\bibfnamefont
  {D.}~\bibnamefont {Hover}}, \bibinfo {author} {\bibfnamefont {M.~E.}\
  \bibnamefont {Schwartz}}, \bibinfo {author} {\bibfnamefont {V.}~\bibnamefont
  {Bolkhovsky}}, \bibinfo {author} {\bibfnamefont {X.}~\bibnamefont {Zhang}},
  \bibinfo {author} {\bibfnamefont {W.~D.}\ \bibnamefont {Oliver}}, \ and\
  \bibinfo {author} {\bibfnamefont {I.}~\bibnamefont {Siddiqi}},\ }\href
  {\doibase 10.1126/science.aaa8525} {\bibfield  {journal} {\bibinfo  {journal}
  {Science}\ }\textbf {\bibinfo {volume} {350}},\ \bibinfo {pages} {307}
  (\bibinfo {year} {2015})}\BibitemShut {NoStop}%
\bibitem [{\citenamefont {Versluis}\ \emph {et~al.}(2017)\citenamefont
  {Versluis}, \citenamefont {Poletto}, \citenamefont {Khammassi}, \citenamefont
  {Tarasinski}, \citenamefont {Haider}, \citenamefont {Michalak}, \citenamefont
  {Bruno}, \citenamefont {Bertels},\ and\ \citenamefont
  {DiCarlo}}]{Versluis17}%
  \BibitemOpen
  \bibfield  {author} {\bibinfo {author} {\bibfnamefont {R.}~\bibnamefont
  {Versluis}}, \bibinfo {author} {\bibfnamefont {S.}~\bibnamefont {Poletto}},
  \bibinfo {author} {\bibfnamefont {N.}~\bibnamefont {Khammassi}}, \bibinfo
  {author} {\bibfnamefont {B.}~\bibnamefont {Tarasinski}}, \bibinfo {author}
  {\bibfnamefont {N.}~\bibnamefont {Haider}}, \bibinfo {author} {\bibfnamefont
  {D.~J.}\ \bibnamefont {Michalak}}, \bibinfo {author} {\bibfnamefont
  {A.}~\bibnamefont {Bruno}}, \bibinfo {author} {\bibfnamefont
  {K.}~\bibnamefont {Bertels}}, \ and\ \bibinfo {author} {\bibfnamefont
  {L.}~\bibnamefont {DiCarlo}},\ }\href {\doibase
  10.1103/PhysRevApplied.8.034021} {\bibfield  {journal} {\bibinfo  {journal}
  {Phys. Rev. Appl.}\ }\textbf {\bibinfo {volume} {8}},\ \bibinfo {pages}
  {034021} (\bibinfo {year} {2017})}\BibitemShut {NoStop}%
\bibitem [{\citenamefont {Johnson}\ \emph {et~al.}(2016)\citenamefont
  {Johnson}, \citenamefont {Ungaretti} \emph {et~al.}}]{QCoDeS16}%
  \BibitemOpen
  \bibfield  {author} {\bibinfo {author} {\bibfnamefont {A.}~\bibnamefont
  {Johnson}}, \bibinfo {author} {\bibfnamefont {G.}~\bibnamefont {Ungaretti}},
  \emph {et~al.},\ }\href {https://github.com/QCoDeS/Qcodes} {\enquote
  {\bibinfo {title} {Q{C}o{D}e{S}},}\ } (\bibinfo {year} {2016})\BibitemShut
  {NoStop}%
\bibitem [{\citenamefont {Rol}\ \emph {et~al.}(2016)\citenamefont {Rol},
  \citenamefont {Dickel}, \citenamefont {Asaad}, \citenamefont {Bultink},
  \citenamefont {Sagastizabal}, \citenamefont {Langford}, \citenamefont
  {de~Lange}, \citenamefont {Dikken}, \citenamefont {Fu}, \citenamefont
  {de~Jong},\ and\ \citenamefont {Luthi}}]{PycQED16}%
  \BibitemOpen
  \bibfield  {author} {\bibinfo {author} {\bibfnamefont {M.~A.}\ \bibnamefont
  {Rol}}, \bibinfo {author} {\bibfnamefont {C.}~\bibnamefont {Dickel}},
  \bibinfo {author} {\bibfnamefont {S.}~\bibnamefont {Asaad}}, \bibinfo
  {author} {\bibfnamefont {C.~C.}\ \bibnamefont {Bultink}}, \bibinfo {author}
  {\bibfnamefont {R.}~\bibnamefont {Sagastizabal}}, \bibinfo {author}
  {\bibfnamefont {N.~K.}\ \bibnamefont {Langford}}, \bibinfo {author}
  {\bibfnamefont {G.}~\bibnamefont {de~Lange}}, \bibinfo {author}
  {\bibfnamefont {B.~C.~S.}\ \bibnamefont {Dikken}}, \bibinfo {author}
  {\bibfnamefont {X.}~\bibnamefont {Fu}}, \bibinfo {author} {\bibfnamefont
  {S.~R.}\ \bibnamefont {de~Jong}}, \ and\ \bibinfo {author} {\bibfnamefont
  {F.}~\bibnamefont {Luthi}},\ }\href {https://doi.org/10.5281/zenodo.160327} {\enquote {\bibinfo {title}
  {Pyc{Q}{E}{D}},}\ } (\bibinfo {year} {2016})\BibitemShut {NoStop}%
\bibitem [{\citenamefont {Fu}\ \emph {et~al.}(2019)\citenamefont {Fu},
  \citenamefont {Riesebos}, \citenamefont {Rol}, \citenamefont {van Straten},
  \citenamefont {van Someren}, \citenamefont {Khammassi}, \citenamefont
  {Ashraf}, \citenamefont {Vermeulen}, \citenamefont {Newsum}, \citenamefont
  {Loh}, \citenamefont {de~Sterke}, \citenamefont {Vlothuizen}, \citenamefont
  {Schouten}, \citenamefont {Almudever}, \citenamefont {DiCarlo},\ and\
  \citenamefont {Bertels}}]{Fu19}%
  \BibitemOpen
  \bibfield  {author} {\bibinfo {author} {\bibfnamefont {X.}~\bibnamefont
  {Fu}}, \bibinfo {author} {\bibfnamefont {L.}~\bibnamefont {Riesebos}},
  \bibinfo {author} {\bibfnamefont {M.~A.}\ \bibnamefont {Rol}}, \bibinfo
  {author} {\bibfnamefont {J.}~\bibnamefont {van Straten}}, \bibinfo {author}
  {\bibfnamefont {J.}~\bibnamefont {van Someren}}, \bibinfo {author}
  {\bibfnamefont {N.}~\bibnamefont {Khammassi}}, \bibinfo {author}
  {\bibfnamefont {I.}~\bibnamefont {Ashraf}}, \bibinfo {author} {\bibfnamefont
  {R.~F.~L.}\ \bibnamefont {Vermeulen}}, \bibinfo {author} {\bibfnamefont
  {V.}~\bibnamefont {Newsum}}, \bibinfo {author} {\bibfnamefont {K.~K.~L.}\
  \bibnamefont {Loh}}, \bibinfo {author} {\bibfnamefont {J.~C.}\ \bibnamefont
  {de~Sterke}}, \bibinfo {author} {\bibfnamefont {W.~J.}\ \bibnamefont
  {Vlothuizen}}, \bibinfo {author} {\bibfnamefont {R.~N.}\ \bibnamefont
  {Schouten}}, \bibinfo {author} {\bibfnamefont {C.~G.}\ \bibnamefont
  {Almudever}}, \bibinfo {author} {\bibfnamefont {L.}~\bibnamefont {DiCarlo}},
  \ and\ \bibinfo {author} {\bibfnamefont {K.}~\bibnamefont {Bertels}},\ }in\
  \href {https://arxiv.org/abs/1808.02449} {\emph {\bibinfo {booktitle}
  {Proceedings of 25th IEEE International Symposium on High-Performance
  Computer Architecture (HPCA)}}}\ (\bibinfo {organization} {IEEE},\ \bibinfo
  {year} {2019})\ pp.\ \bibinfo {pages} {224--237}\BibitemShut {NoStop}%
\bibitem [{\citenamefont {Saira}\ \emph {et~al.}(2014)\citenamefont {Saira},
  \citenamefont {Groen}, \citenamefont {Cramer}, \citenamefont {Meretska},
  \citenamefont {de~Lange},\ and\ \citenamefont {DiCarlo}}]{Saira14}%
  \BibitemOpen
  \bibfield  {author} {\bibinfo {author} {\bibfnamefont {O.-P.}\ \bibnamefont
  {Saira}}, \bibinfo {author} {\bibfnamefont {J.~P.}\ \bibnamefont {Groen}},
  \bibinfo {author} {\bibfnamefont {J.}~\bibnamefont {Cramer}}, \bibinfo
  {author} {\bibfnamefont {M.}~\bibnamefont {Meretska}}, \bibinfo {author}
  {\bibfnamefont {G.}~\bibnamefont {de~Lange}}, \ and\ \bibinfo {author}
  {\bibfnamefont {L.}~\bibnamefont {DiCarlo}},\ }\href
  {https://journals.aps.org/prl/abstract/10.1103/PhysRevLett.112.070502}
  {\bibfield  {journal} {\bibinfo  {journal} {Phys. Rev. Lett.}\ }\textbf
  {\bibinfo {volume} {112}},\ \bibinfo {pages} {070502} (\bibinfo {year}
  {2014})}\BibitemShut {NoStop}%
\bibitem [{\citenamefont {Chow}\ \emph {et~al.}(2012)\citenamefont {Chow},
  \citenamefont {Gambetta}, \citenamefont {C\'orcoles}, \citenamefont {Merkel},
  \citenamefont {Smolin}, \citenamefont {Rigetti}, \citenamefont {Poletto},
  \citenamefont {Keefe}, \citenamefont {Rothwell}, \citenamefont {Rozen},
  \citenamefont {Ketchen},\ and\ \citenamefont {Steffen}}]{Chow12}%
  \BibitemOpen
  \bibfield  {author} {\bibinfo {author} {\bibfnamefont {J.~M.}\ \bibnamefont
  {Chow}}, \bibinfo {author} {\bibfnamefont {J.~M.}\ \bibnamefont {Gambetta}},
  \bibinfo {author} {\bibfnamefont {A.~D.}\ \bibnamefont {C\'orcoles}},
  \bibinfo {author} {\bibfnamefont {S.~T.}\ \bibnamefont {Merkel}}, \bibinfo
  {author} {\bibfnamefont {J.~A.}\ \bibnamefont {Smolin}}, \bibinfo {author}
  {\bibfnamefont {C.}~\bibnamefont {Rigetti}}, \bibinfo {author} {\bibfnamefont
  {S.}~\bibnamefont {Poletto}}, \bibinfo {author} {\bibfnamefont {G.~A.}\
  \bibnamefont {Keefe}}, \bibinfo {author} {\bibfnamefont {M.~B.}\ \bibnamefont
  {Rothwell}}, \bibinfo {author} {\bibfnamefont {J.~R.}\ \bibnamefont {Rozen}},
  \bibinfo {author} {\bibfnamefont {M.~B.}\ \bibnamefont {Ketchen}}, \ and\
  \bibinfo {author} {\bibfnamefont {M.}~\bibnamefont {Steffen}},\ }\href
  {\doibase 10.1103/PhysRevLett.109.060501} {\bibfield  {journal} {\bibinfo
  {journal} {Phys. Rev. Lett.}\ }\textbf {\bibinfo {volume} {109}},\ \bibinfo
  {pages} {060501} (\bibinfo {year} {2012})}\BibitemShut {NoStop}%
\bibitem [{\citenamefont {Greenbaum}(2015)}]{Greenbaum}%
  \BibitemOpen
  \bibfield  {author} {\bibinfo {author} {\bibfnamefont {D.}~\bibnamefont
  {Greenbaum}},\ }\href {https://arxiv.org/abs/1509.02921} {\bibfield
  {journal} {\bibinfo  {journal} {arXiv:1509.02921}\ } (\bibinfo {year}
  {2015})}\BibitemShut {NoStop}%
\end{thebibliography}
\end{document}